\documentclass[amsmath,amssymb,pre]{revtex4}
\usepackage[]{graphicx} 
\usepackage{latexsym,amsmath,amsfonts,amssymb,bm,empheq}
\usepackage{multirow,booktabs}
\usepackage{threeparttable}
\usepackage{here}

\begin{document}
\title{Pole Analysis of the Inter-Replica Correlation Function in a Two-Replica System as a Binary Mixture: Mean Overlap in the Cluster Glass Phase}

\author{Hiroshi Frusawa}
\email{frusawa.hiroshi@kochi-tech.ac.jp}
\affiliation{Laboratory of Statistical Physics, Kochi University of Technology, Tosa-Yamada, Kochi 782-8502, Japan.}
\date{\today}
\begin{abstract}
To investigate the cluster glass phase of ultrasoft particles, we examine an annealed two-replica system endowed with an attractive inter-replica field similar to that of a binary symmetric electrolyte. Leveraging this analogy, we conduct pole analysis on the total correlation functions in the two-replica system where the inter-replica field will eventually be switched off. By synthesizing discussions grounded in the pole analysis with a hierarchical view of the free-energy landscape, we derive an analytical form of the mean overlap between two replicas within the mean field approximation of the Gaussian core model. This formula elucidates novel numerical findings observed in the cluster glass phase.
\end{abstract}
\maketitle

\section{Introduction}\label{sec intro}
Ultrasoft particles that may overlap and deform offer a powerful model for understanding complex systems such as polymers, dendrimers, and two-dimensional bosons~\cite{likos rev1,likos rev2,van hecke,nature,vortex}.
The theoretical and simulation results on the soft-core models have revealed new and surprising features, including clustering phenomena and reentrant phase behaviors \cite{likos rev1,likos rev2,van hecke,nature,vortex,ikeda1,ikeda plus,ikeda2,coslo1,coslo2,ikeda3,miyazaki1,miyazaki2,cl1,cl2,cl3,cl4,cl5,cl6,cl7,cl8,likos g1,likos g2}.
For example, the generalized exponential model (GEM) adopts an ultrasoft potential $v(r)$ in units of $k_BT$ as follows:
\begin{eqnarray}
v(r)=\widetilde{\epsilon}\exp\left\{-r^s\right\},
\label{gem potential}
\end{eqnarray}
where interparticle distance $r$ is measured in units of a characteristic length $d$, $\widetilde{\epsilon}$ corresponds to a dimensionless interaction strength at zero separation, and the potential profile is characterized by a softness exponent $s$ \cite{ikeda1,ikeda plus,ikeda2,coslo1,coslo2,ikeda3,miyazaki1,miyazaki2,cl1,cl2,cl3,cl4,cl5,cl6,cl7,cl8,likos g1,likos g2}.

{Especially for $s>2$ of the GEM, or the so-called $Q^{\pm}$ class, the non-positive definiteness of the Fourier transform $v(k)$ of $v(r)$ leads to the formation of equilibrium clusters in the liquid, crystalline, and glass phases, as confirmed by theoretical and simulation \mbox{studies~\cite{ikeda1,ikeda plus,ikeda2,coslo1,coslo2,ikeda3,miyazaki1,miyazaki2,cl1,cl2,cl3,cl4,cl5,cl6,cl7,cl8,likos g1,likos g2}.}
Meanwhile, the complementary class ($Q^{+}$ class) of the GEM for $s\leq 2$ does not form clusters in equilibrium due to the positive definite Fourier transform \mbox{(i.e., $v(k)>0$) \cite{ikeda1,ikeda plus,ikeda2,coslo1,coslo2,ikeda3,miyazaki1,miyazaki2,cl1,cl2,cl3,cl4,cl5,cl6,cl7,cl8,likos g1,likos g2}}.
Recently, however, the glass phase of the GEM at $s=2$, or the Gaussian core model (GCM), has been investigated using two copies of replicated systems (two-replica system), thereby demonstrating the occurrence of local aggregates or out-of-equilibrium clusters in the glass phase \cite{likos g1,likos g2}.}

We are concerned with the cluster-forming glassy states of soft-core models, particularly the GEM of ultrasoft particles interacting via $v(r)$ irrespective of whether $v(r)$ given by Equation (\ref{gem potential}) belongs to the $Q^{\pm}$ class ($s>2$) or the $Q^{+}$ class ($s\leq 2$).

The cluster-forming states tend to emerge at higher densities where the supercooled ultrasoft systems behave as mean-field fluids \cite{ikeda1,ikeda plus,ikeda2,coslo1,coslo2,ikeda3,miyazaki1,miyazaki2,cl1,cl2,cl3,cl4,cl5,cl6,cl7,cl8,likos g1,likos g2}.
Correspondingly, experimental and simulation studies on single-particle dynamics in the cluster glasses have revealed the following hierarchical feature, other than the cage-hopping dynamic characteristic of non-cluster-forming supercooled liquids:
while the cooperative dynamics are ascribed to the freezing of clusters' center of mass, the single-particle dynamics include inter-cluster hopping and intra-cluster fluctuations \cite{likos g2,d1,d2,d3,d4,d5}.
The blurring of cage-hopping dynamics in the high-density region is similar to that of the short-time dynamics in the marginal glass phase~\cite{ma1,ma2,ma3,ma4,ma5,ma6,ma7}.
While a stable glass has a plateau in the mean squared displacement (MSD) that decreases with increasing density, a marginal glass exhibits a logarithmic growth of the MSD with lag time \cite{ma1,ma2,ma3,ma4,ma5,ma6,ma7}.
The former represents the crossover from ballistic motion to caged behavior, whereas the latter indicates that particles leave small cages to find themselves in slightly larger cages: cages themselves are ever-shifting \cite{ma1,ma2,ma3,ma4,ma5,ma6,ma7}.

For comparison, it is helpful to see the stable and marginal glasses in terms of the free-energy landscape (FEL) \cite{ma1,ma2,ma3,ma4,ma5,ma6,ma7}.
On the one hand, the FEL in the stable glass state consists of many smooth basins separated by energy barriers.
The single-particle dynamics in the stable glass include rattling inside an arrested cage and cage-to-cage hopping, which are interpreted as vibrational excitations around each basin and jumps between different basins, respectively.
The marginal glass, on the other hand, provides infinitesimally different metastable states, reflecting the hierarchical FEL where the basin splits into a fractal hierarchy of \mbox{sub-basins \cite{ma1,ma2,ma3,ma4,ma5,ma6,ma7}.}
As a result, even an infinitesimal perturbation can trigger a transition between adjacent similar metastable states. 
The above dynamical similarity between the cluster-forming and marginal glasses suggests that it is relevant to consider the hierarchical FEL \cite{ma1,ma2,ma3,ma4,ma5,ma6,ma7} when addressing the two-scale fluctuations in cluster-forming glasses.

Here, we develop a theoretical framework for the two-replica system without a quenched disorder, or the annealed two-replica system \cite{qu1,qu2,qu3,qu4,qu5,qu6,qu7,an1,an2,an3,an4,an5,an6,an7}, in the cluster-forming glassy states.
Our focus is on the pole analysis \cite{evans1,evans2,evans3,evans4,evans5,evans6} of density--density correlation functions considering the hierarchical FEL.
The previous results demonstrate the usefulness of the pole analysis when investigating fluctuations separately by length scales.
For example, the pole analysis method applies to a fluid mixture with competing interactions that cause cluster formation, thereby detecting the change in density--density correlation functions from monotonic to damped oscillatory decay \cite{evans1,evans2,evans3,evans4,evans5,evans6,com1,com2,com3,com4,com5,com6,com7}; the fluid mixture undergoes continuous transitions to a state characterized by periodic concentration fluctuations (microphase separation).
A combined approach of the pole analysis and FEL would thus be relevant to investigate the annealed two-replica system \cite{an1,an2,an3,an4,an5,an6,an7} in the cluster-forming glassy states where the cage-to-cage hopping is obscured \cite{likos g2,d1,d2,d3,d4,d5}.

This paper is organized as follows.
In Section \ref{sec basic}, we introduce mean overlap as an index by which to measure similarity between two replicas and express it using total correlation functions (TCFs).
In Section \ref{sec general}, we conduct the pole analysis of TCFs, providing novel results that generally apply to two-replica systems in glassy states.
Section \ref{sec tcf} demonstrates that the pole-analysis-based discussions allow us to determine the TCFs without solving the Ornstein--Zernike (OZ) equations.
In Section \ref{sec insights}, we consider the FEL in the cluster glass phase for revealing the underlying physics of TCFs obtained from the pole-analysis-based discussions.
In Section \ref{sec mean}, we assess the validity of the pole-analysis-based results, focusing on the GCM in the mean field approximation.
Section \ref{sec concluding} makes some concluding remarks.

\section{Two-Replica System: Basic Formulation}\label{sec basic}
This section introduces mean overlap and TCF, characteristic quantities of the two-replica system \cite{qu1,qu2,qu3,qu4,qu5,qu6,qu7,an1,an2,an3,an4,an5,an6,an7}.
First, in Section \ref{sub overlap}, we define the mean overlap using a density correlation function as an index by which to measure the similarity between two replicas. Next, in Section \ref{sub binary}, based on the analogy with symmetric binary electrolytes, two replicas are characterized by the average density, the difference density (corresponding to the charge density in electrolytes), and the associated TCFs that provide intra- and inter-replica ones. Finally, in Section \ref{sub oz}, we present the OZ equations for symmetric binary mixtures \cite{an3,an4,an5,an6,an7,evans6}, which allows us to express TCFs of average and difference densities using the direct correlation functions (DCFs).

\subsection{Mean Overlap}\label{sub overlap}
Let $\widehat{\rho}_{\alpha}(\bm{x})=\sum_{i=1}^N\delta\left[\bm{x}-\bm{r}_{\alpha}^i\right]$ be a microscopic density in replica $\alpha\in\{1,\,2,\cdots,n\}$ of the $N$-particle system with the volume $V$ when considering the $i$th-particle located at $\bm{r}_{\alpha}^i\,(1\leq i\leq N)$.
The mean overlap $Q_{\alpha\beta}$ represents the degree of similarity between configurations of replica $\alpha$ and replica $\beta$.
We define $Q_{\alpha\beta}$ as
\begin{flalign}
Q_{\alpha\beta}&=\frac{d^6}{N}\iint d\bm{x}d\bm{y}\,
\Theta\left(l-r\right)\left<\widehat{G}_{\alpha\beta}(r)\right>,
\label{mean overlap}
\\
\widehat{G}_{\alpha\beta}(r)&=\widehat{\rho}_{\alpha}(\bm{x})\,\widehat{\rho}_{\beta}(\bm{y})
\qquad(\alpha\neq\beta),
\label{inst overlap}
\end{flalign}
where $r\equiv |\bm{x}-\bm{y}|$, the Heaviside function $\Theta\left(l-r\right)$ is used to exclude the region of $r>l$, and the brackets denote the statistical average over the realizations of particle configurations~\cite{qu1,qu2,qu3,qu4,qu5,qu6,qu7,an1,an2,an3,an4,an5,an6,an7}.

While the replica trick for structural glasses eventually takes the limit of either $n\rightarrow 0$ or $n\rightarrow 1$ for the number of replicas \cite{qu7}, this paper is concerned with the mean overlap at \mbox{$n=2$.}
The two-replica overlap can also be an indicator of glassy states by switching on and off the inter-replica attractive field, similar to the role of a magnetic field in studying magnetization.
Introducing an external field $-\widetilde{\epsilon}_{12}$ conjugate to the mean overlap, Equation~(\ref{mean overlap}) allows us to write the inter-replica attractive energy $U_{\mathrm{att}}$ in units of $k_BT$ as
\begin{eqnarray}
U_{\mathrm{att}}=-N\widetilde{\epsilon}_{12}Q_{12}
\end{eqnarray}
for the two-replica system.
The attractive inter-replica field $-\widetilde{\epsilon}_{12}$ plays different roles in the quenched and annealed systems.
While the quenched system in replica 1 is studied in the presence of a quenched disorder created by a reference configuration of replica 2 drawn from the equilibrium distribution \cite{qu1,qu2,qu3,qu4,qu5,qu6,qu7}, the annealed system considers configurations of two replicas fluctuating simultaneously \cite{an1,an2,an3,an4,an5,an6,an7}:
$-\widetilde{\epsilon}_{12}$ acts on one of the replicas and both replicas in the quenched and annealed dynamics, respectively.

In the annealed system \cite{an1,an2,an3,an4,an5,an6,an7}, the two-replica system can simply be regarded as a binary symmetric electrolyte \cite{evans6}.
This analogy allows us to predict that complexation between particles in replicas 1 and 2 occurs at sufficiently low temperatures due to the inter-replica attraction energy $-N\widetilde{\epsilon}_{12}Q_{12}$.
It has also been demonstrated that such complexes remain stable in the glass phase even when $\widetilde{\epsilon}_{12}$ is progressively lowered from the initial value to zero (i.e., $\widetilde{\epsilon}_{12}\rightarrow 0$), providing a non-trivial value of $Q_{12}$ that is larger than the random overlap $Q_r$ evaluated at $\left<\widehat{G}_{\alpha\beta}(r)\right>=\overline{\rho}^2$ using the uniform density $\overline{\rho}=N/V$.
It follows from Equation (\ref{mean overlap}) that
\begin{flalign}
\label{random}
Q_r&=\frac{4\pi}{3}\phi\,(ld)^3,\\
\label{phi}
\phi&=\overline{\rho}\,d^3,
\end{flalign}
where $\phi$ denotes the volume fraction \cite{ikeda1,ikeda plus,ikeda2,coslo1,coslo2,ikeda3,miyazaki1,miyazaki2,cl1,cl2,cl3,cl4,cl5,cl6,cl7,cl8,likos g1,likos g2,an1,an2,an3,an4,an5,an6,an7}.

\subsection{Two Replicas as a Binary Mixture: Density--Density Correlations}\label{sub binary}
In the annealed two-replica system, it is natural to introduce another set of coarse-grained density variables related to the sum and difference, similar to a symmetric electrolyte \cite{evans6}:
\begin{eqnarray}
\widehat{\rho}_+(\bm{r})=\frac{1}{2}\left\{\widehat{\rho}_1(\bm{r})+\widehat{\rho}_2(\bm{r})\right\},
\label{plus density}
\\
\widehat{\rho}_-(\bm{r})=\frac{1}{2}\left\{\widehat{\rho}_1(\bm{r})-\widehat{\rho}_2(\bm{r})\right\},
\label{minus density}
\end{eqnarray}
which will be called $p$-density and $n$-density, respectively, after the subscripts of positive and negative signs.
As seen from the above definitions, the mean $p$-density $\left<\widehat{\rho}_+(\bm{r})\right>$ corresponds to the average density of two replicas, whereas the mean $n$-density $\left<\widehat{\rho}_-(\bm{r})\right>$ is a measure of local charge density in analogy with symmetric electrolytes.
Accordingly, the uniform distribution results in $\left<\widehat{\rho}_+(\bm{r})\right>=\overline{\rho}$ and $\left<\widehat{\rho}_-(\bm{r})\right>=0$.

We investigate the binary cluster glasses in the limit of $\widetilde{\epsilon}_{12}\rightarrow 0$ using the $p$-$p$ and $n$-$n$ TCFs.
Given that
\begin{eqnarray}
\left<\widehat{G}_{\alpha\beta}(r)\right>
=\overline{\rho}^2\left\{1+h_{\alpha\beta}(r)\right\}
\label{r total}
\end{eqnarray}
for the TCF $h_{\alpha\beta}(r)$ between replica $\alpha$ and replica $\beta$, we have
\begin{flalign}
\left<\widehat{G}_{++}(r)\right>
&=\overline{\rho}^2\left\{1+h_{+}(r)\right\},
\label{p total}\\
\left<\widehat{G}_{--}(r)\right>
&=\overline{\rho}^2h_{-}(r),
\label{n total}    
\end{flalign}
respectively, where
\begin{flalign}
\widehat{G}_{++}(r)&=\frac{1}{4}\left\{\widehat{G}_{11}(r)+2\widehat{G}_{12}(r)+\widehat{G}_{22}(r)\right\},
\label{g++}\\
\widehat{G}_{--}(r)&=\frac{1}{4}\left\{\widehat{G}_{11}(r)-2\widehat{G}_{12}(r)+\widehat{G}_{22}(r)\right\},
\label{g--}
\end{flalign}
{due to Equations (\ref{plus density}) and (\ref{minus density}), and the TCFs $h_{+}(r)$ and $h_{-}(r)$ are defined by Equations (\ref{p total}) and~(\ref{n total}).}

The symmetric system also allows us to introduce the intra- and inter-replica TCFs, $h(r)$ and $\widetilde{h}(r)$, given by
\begin{flalign}
h(r)=h_{11}(r)=h_{22}(r),
\label{def no tilde tcf}\\
\widetilde{h}(r)=h_{12}(r)=h_{21}(r),
\label{def tilde tcf}
\end{flalign}
respectively.

It follows from Equations (\ref{r total})--(\ref{def tilde tcf}) that
\begin{flalign}
\label{h11 pm}
h(r)=h_{+}(r)+h_{-}(r),\\
\label{h12 pm}
\widetilde{h}(r)=h_{+}(r)-h_{-}(r).
\end{flalign}
Equation (\ref{h12 pm}) implies that the inter-replica TCF $\widetilde{h}(r)$ has a non-zero value when there is a difference between the $p$-$p$ and $n$-$n$ correlations even without inter-replica attractive interactions (i.e., $\widetilde{\epsilon}_{12}=0$).
Plugging Equations (\ref{r total}) and (\ref{def tilde tcf}) into Equation (\ref{mean overlap}), we have
\begin{flalign}
\label{meanq1}
Q_{12}=Q_r+
4\pi\phi
\int_0^{l}
\>dr\,r^2\widetilde{h}(r).
\end{flalign}
The main purpose of this paper is to calculate the mean overlap $Q_{12}$ using the inter-replica TCF $\widetilde{h}(r)$ in Equation (\ref{meanq1}).

\subsection{The Ornstein--Zernike Equations in Symmetric Binary Mixtures}\label{sub oz}
The OZ equations in Fourier space relate the TCFs to the intra- and inter-replica DCFs denoted by $c(k)$ and $\widetilde{c}(k)$, respectively, as follows \cite{an3,an4,an5,an6,an7}:
\begin{flalign}
\label{oz1}
h(k)&=c(k)+\overline{\rho}\left\{
c(k)h(k)+\widetilde{c}(k)\widetilde{h}(k)\right\},
\\
\label{oz2}
\widetilde{h}(k)&=\widetilde{c}(k)+\overline{\rho}\left\{
c(k)\widetilde{h}(k)+\widetilde{c}(k)h(k)\right\},
\end{flalign}
which apply to the two-replica system at $n=2$. 
Meanwhile, Equations (\ref{h11 pm}) and (\ref{h12 pm}) read
\begin{flalign}
h_{+}(k)
=\frac{1}{2}\left\{
h(k)+\widetilde{h}(k)
\right\},
\label{p tcf}\\
h_{-}(k)
=\frac{1}{2}\left\{
h(k)-\widetilde{h}(k)
\right\}.
\label{n tcf}
\end{flalign}
Substituting Equations (\ref{oz1}) and (\ref{oz2}) into the right-hand sides (RHSs) of Equations (\ref{p tcf}) and (\ref{n tcf}), we have
\begin{flalign}
h_{+}(k)&=\frac{1}{2}\left\{
\frac{c_{+}(k)}{1-\overline{\rho}c_{+}(k)}
\right\},
\label{hc+}\\
h_{-}(k)&=\frac{1}{2}\left\{
\frac{c_{-}(k)}{1-\overline{\rho}c_{-}(k)}
\right\},
\label{hc-}
\end{flalign}
when defining that
\begin{flalign}
c_+(k)=c(k)+\widetilde{c}(k),
\label{def p dcf}\\
c_-(k)=c(k)-\widetilde{c}(k).
\label{def n dcf}
\end{flalign}
We will conduct the pole analysis of Equations (\ref{hc+}) and (\ref{hc-}).

\section{Pole Analysis: General Results}\label{sec general}
Conducting the pole analysis \cite{evans1,evans2,evans3,evans4,evans5,evans6} of Equations (\ref{hc+}) and (\ref{hc-}), we can find novel results that generally apply to two-replica systems in glassy states.
In Section \ref{sub mix}, we obtain the pole equations for glassy states, revealing new findings of density fluctuations.
In Section \ref{sub req}, we present requirements for the inter-replica TCF in real and Fourier spaces, based on physical considerations.

\subsection{Pole Equations}\label{sub mix}
Considering that the real-space representation $c_{\sigma}(r)$ ($\sigma=+$ or $-$) of the DCF decays faster than a power law for finite-ranged or exponentially decaying interaction potentials, the TCF given by either Equation (\ref{hc+}) or Equation (\ref{hc-}) have poles at complex wavenumbers, $k_m^{\sigma}=a_m^{\sigma}+ib_m^{\sigma}\,(m=1,\,2,\cdots)$ \cite{evans1,evans2,evans3,evans4,evans5,evans6}.
The integrated representation of pole equations is
\begin{flalign}
1-\overline{\rho}c_{\sigma}(k_m^{\sigma})
&=1-\overline{\rho}\left\{
c(k_m^{\sigma})
+\sigma\widetilde{c}(k_m^{\sigma})\right\}
\nonumber\\
&=0,
\label{dcf pole}
\end{flalign}
providing the $m$th pole of $h_{\sigma}(k)$ given by either Equation (\ref{hc+}) for $\sigma=+$ or Equation (\ref{hc-}) for $\sigma=-$.

It follows from the OZ Equation (\ref{oz2}) that the inter-replica DCF reads
\begin{eqnarray}
\label{inter dcf}
\overline{\rho}\widetilde{c}(k)=\left\{1-\overline{\rho}c(k)\right\}
\frac{\overline{\rho}\widetilde{h}(k)}{1+\overline{\rho}h(k)}.
\end{eqnarray}
Plugging Equation (\ref{inter dcf}) into the RHS of the first line in Equation (\ref{dcf pole}), the $m$th pole equation becomes
\begin{flalign}
&\left\{1-\overline{\rho}c(k_m^{\sigma})\right\}
\left\{1-
\frac{\sigma\overline{\rho}\widetilde{h}(k_m^{\sigma})}{1+\overline{\rho}h(k_m^{\sigma})}
\right\}
\nonumber\\
&\qquad
=\frac{1-\overline{\rho}c(k_m^{\sigma})}{1+\overline{\rho}h(k_m^{\sigma})}
\left[1+\overline{\rho}\left\{
h(k_m^{\sigma})-\sigma \widetilde{h}(k_m^{\sigma})
\right\}\right]
\nonumber\\
&\qquad
=0.
\label{dcf pole2}
\end{flalign}
In the supercooled liquid state, Equation (\ref{dcf pole2}) reduces to 
\begin{eqnarray}
1-\overline{\rho}c(k_m)=0,
\label{dcf pole3}
\end{eqnarray}
because of $\widetilde{h}(k)=0$;
we find from the Frourier transform of Equation (\ref{h12 pm}) that $k_m=k_m^+=k_m^-$ is a sufficient condition for $\widetilde{h}(k)=0$.

Meanwhile, in a glassy state where $\widetilde{h}(k)\neq 0$, Equation (\ref{dcf pole2}) provides another equation,
\begin{flalign}
1+\overline{\rho}\left\{
h(k_m^{\sigma})-\sigma\widetilde{h}(k_m^{\sigma})\right\}=0,
\label{main pole}
\end{flalign}
in addition to Equation (\ref{dcf pole3}), or
\begin{eqnarray}
\widetilde{c}(k_m)=0
\label{dcf pole4}
\end{eqnarray}
due to the original pole Equation (\ref{dcf pole}).
Equations (\ref{hc+})--(\ref{dcf pole4}) indicate the following:
\begin{itemize}
\item Combining Equations (\ref{hc+}), (\ref{hc-}), (\ref{dcf pole}), and (\ref{main pole}), it becomes apparent that a sign reversal occurs in glassy states:
the pole equation for $c_{\sigma}(k)$ becomes that for $h_{-\sigma}(k)$.
\item Equations (\ref{dcf pole3})--(\ref{dcf pole4}) imply that there are two kinds of poles for TCFs, $h_+(k)$ and $h_-(k)$, in glassy states: one type is constituted by some of the same poles as those in the liquid state, and the other by the poles satisfying Equation (\ref{main pole}).
\end{itemize}

It is instructive to examine the underlying physics of the pole equation expressed by Equation (\ref{main pole}).
To do so, we introduce the structure factors as follows:
\begin{flalign}
S(k)&=N^{-1}\left<\widehat{\rho}_{\alpha}(k)\widehat{\rho}_{\alpha}(-k)\right>,
\label{def s}\\
\widetilde{S}(k)&=N^{-1}\left<\widehat{\rho}_{\alpha}(k)\widehat{\rho}_{\beta}(-k)\right>
\quad(\alpha\neq\beta),
\label{def tilde s}\\
S_{\sigma}(k)&=N^{-1}\left<\widehat{\rho}_{\sigma}(k)\widehat{\rho}_{\sigma}(-k)\right>,
\label{def s sigma}
\end{flalign}
for $1\leq \alpha,\,\beta\leq 2$, and $\sigma=+$ or $-$.
Considering the density relations in Equations (\ref{plus density}) and~(\ref{minus density}), Equations (\ref{def s})--(\ref{def s sigma}) yield
\begin{flalign}
S_{\sigma}(k)=\frac{1}{2}\left\{
S(k)+\sigma\widetilde{S}(k)
\right\},
\label{s relation}
\end{flalign}
which becomes
\begin{flalign}
S_{\sigma}(k)&=\frac{1}{2}\left\{
1+\overline{\rho}h(k)+\sigma\overline{\rho}\widetilde{h}(k)
\right\}\nonumber\\
&=\frac{1}{2}+\overline{\rho}h_{\sigma}(k),
\label{s relation2}
\end{flalign}
noting that $S(k)=1+\overline{\rho}h(k)$ and $\widetilde{S}(k)=\overline{\rho}\widetilde{h}(k)$ as well as the definition of $h_{\sigma}(k)$ in Equation (\ref{p tcf}) or Equation (\ref{n tcf}).
Combining Equations (\ref{main pole}) and (\ref{s relation2}), we have
\begin{flalign}
\label{main2 pole p}
&S_{-}(k_m^+)=0,\\
\label{main2 pole n}
&S_{+}(k_m^-)=0.
\end{flalign}
Sign reversals appear in Equations (\ref{main2 pole p}) and (\ref{main2 pole n}), as mentioned above:
$S_{\sigma}(k)$ as a function of complex wave number $k$ equals zero at a complex wavenumber of $k_m^{-\sigma}$.

Among our main theoretical findings are Equations (\ref{main2 pole p}) and (\ref{main2 pole n}).
The mixed relationship implies the following characteristics of glassy states:
\begin{itemize}
\item {\itshape{Opposite phenomena}:
the enhancement of $p$-density fluctuations at $k_m^+$ leads to the suppression of $n$-density fluctuations and vice versa, which creates the inter-replica correlations.}
\item {\itshape{Exclusive poles}}: the above reverse trend in density fluctuations indicates that there are unique poles $k_m^{\sigma}$ that $h_+(k)$ and $h_-(k)$ do not share with each other.
\end{itemize}
Section \ref{sub two lengths} will relate the exclusive poles to characteristic lengths of cluster glasses. 

\subsection{Two Requirements}\label{sub req}
The poles determined by Equations (\ref{main2 pole p}) and (\ref{main2 pole n}) allow us to perform the 3D Fourier transform of Equations (\ref{hc+}) and (\ref{hc-}) using contour integral along an infinite radius semicircle in the complex upper half-plane.
Remembering that $k_m^{\sigma}=a_m^{\sigma}+ib_m^{\sigma}\,(m=1,\,2,\cdots)$ at poles, the pole analysis provides \cite{evans1,evans2,evans3}
\begin{flalign}
rh_{\sigma}(r)=
\sum_{m\geq1}A_m^{\sigma}e^{-b_m^{\sigma}r}\cos\left(
a_m^{\sigma}r+\theta_m^{\sigma}
\right),
\label{general tcf}
\end{flalign}
indicating that $(b_m^{\sigma})^{-1}$ corresponds to the decay length.
We use the pole series in order of decreasing decay lengths: $b_m^{\sigma}<b_{m+1}^{\sigma}$.
The contribution from the $m$th pole in Equation (\ref{general tcf}) will be referred to as the $m$th mode.

Let $h^{\mathrm{liq}}_{\sigma}(r)$ and $h^{\mathrm{liq}}(r)$ denote TCFs of the supercooled liquid before vitrification.
\mbox{Equation~(\ref{h12 pm})} implies that there is no difference between $h^{\mathrm{liq}}_{+}(r)$ and $h^{\mathrm{liq}}_{-}(r)$ because of the absence of the inter-replica TCF in the supercooled liquid state.
Therefore, Equation~(\ref{general tcf}) reduces to
\begin{flalign}
rh^{\mathrm{liq}}_+(r)=
rh^{\mathrm{liq}}_-(r)=
\sum_{m\geq1}A_me^{-b_mr}\cos\left(a_mr+\theta_m\right), 
\label{liquid tcf1}
\end{flalign}
for $k_m=a_m+ib_m$, satisfying Equation (\ref{dcf pole3}).
Combining Equations (\ref{h11 pm}) and (\ref{liquid tcf1}), the intra-replica TCF $h^{\mathrm{liq}}(r)$ in the supercooled liquid state becomes
\begin{flalign}
\label{eqm}
rh^{\mathrm{liq}}(r)=\sum_{m\geq1}2A_me^{-b_mr}\cos\left(a_mr+\theta_m\right),
\end{flalign}
where the $m$th mode is determined by the pole Equation (\ref{dcf pole3}).

It has been confirmed that the above type of expansion form based on the pole analysis correctly yields the correlation functions in the liquid state \cite{evans1,evans2,evans3,evans4,evans5,evans6}.
It has also been found that liquids and glasses are indistinguishable from each other in terms of two-body density correlation:
there is little structural change accompanying the slowdown of the dynamics of supercooled liquids.

The above structural feature leads to the requirement [RR] on glassy states in real~space:
\begin{flalign}
[\mathrm{RR}]\quad
h(r)=h_{+}(r)+h_{-}(r)= h^{\mathrm{liq}}(r),
\nonumber
\end{flalign}
similar to $h_{+}^{\mathrm{liq}}(r)+h_{-}^{\mathrm{liq}}(r)= h^{\mathrm{liq}}(r)$ in Equation (\ref{eqm}), while producing
\begin{flalign}
\widetilde{h}(r)
=h_{+}(r)-h_{-}(r)\neq 0
\label{rr plus}
\end{flalign}
by definition of glassy states.
The real-space requirement [RR] with Equations (\ref{eqm}) and~ (\ref{rr plus}) states that the sum of $h_{+}(r)$ and $h_{-}(r)$ in glassy states is constituted by the liquid-state modes represented by $k_m$, despite making a difference between $h_{+}(r)$ and $h_{-}(r)$.

In Fourier space, on the other hand, Equations (\ref{dcf pole3}) and (\ref{main pole}), as well as the real-space requirement [RR], imply that the glassy poles given by Equation (\ref{main pole}) must satisfy \mbox{Equations (\ref{dcf pole3}) and (\ref{dcf pole4})} simultaneously;
otherwise, the exclusive modes other than the liquid-state ones necessarily remain by adding $h_{+}(r)$ and $h_{-}(r)$ against the real-space requirement [RR] because of the exclusive poles explained at the end of Section \ref{sub mix}. 
The constraint [RR] therefore applies only to solutions that satisfy Equations (\ref{dcf pole3}) to (\ref{dcf pole4}) simultaneously. 
Thus, the Fourier space requirement [FR], or another expression of the real-space one [RR], is that the pole Equations (\ref{main2 pole p}) and (\ref{main2 pole n}) for glassy states must hold at a subset of $k_m$ satisfying Equation (\ref{dcf pole3}), or liquid-state modes:
\begin{flalign}
[\mathrm{FR}]\quad
S_{-}(k_{\nu})=0,
\quad S_{+}(k_{\nu'})=0,
\nonumber
\end{flalign}
for $\nu,\,\nu'\subset m$ with $\nu\neq\nu'$.
It is to be remembered from Equations (\ref{main2 pole p}) and (\ref{main2 pole n}) that $k_{\nu}$ corresponds to the pole of $h_{+}(k)$ and $k_{\nu'}$ to that of $h_-(k)$.

\section{Inter-Replica Total Correlation Function}\label{sec tcf}
This section demonstrates that the pole-analysis-based discussions allow us to determine the inter-replica TCF $\widetilde{h}(r)$ without finding out-of-equilibrium solutions to the OZ Equations (\ref{oz1}) and (\ref{oz2}).
In Section \ref{sub general}, we obtain a general form of the inter-replica TCF in real space by combining the requirements [RR] and [FR]. Furthermore, we confirm that the pole Equations (\ref{dcf pole3}) to (\ref{dcf pole4}) yield the Fourier transform of the general expression.
In \mbox{Section \ref{sub two mode}}, we propose a two-mode model considering the two-scale dynamics in cluster~glasses.

\subsection{General Form}\label{sub general}
The two requirements, [RR] and [FR], provide
\begin{flalign}
r\widetilde{h}(r)=
\sum_{\nu}2A_{\nu}e^{-b_{\nu}r}\cos\left(a_{\nu}r+\theta_{\nu}\right)
-\sum_{\nu'}2A_{\nu'}e^{-b_{\nu'}r}\cos\left(a_{\nu'}r+\theta_{\nu'}\right).
\label{h12 simple}
\end{flalign}
Equations (\ref{eqm}) and (\ref{h12 simple}) satisfy Equations (\ref{h11 pm}) and (\ref{h12 pm}) by taking the following forms:
\begin{flalign}
rh_+(r)=&\sum_{m}A_me^{-b_mr}\cos\left(a_mr+\theta_m\right)
+\frac{r\widetilde{h}(r)}{2},
\label{h+ simple}\\
rh_-(r)=&\sum_{m}A_me^{-b_mr}\cos\left(a_mr+\theta_m\right)
-\frac{r\widetilde{h}(r)}{2}.
\label{h- simple}
\end{flalign}
Equation (\ref{h12 simple}) corresponds to a general expression of the inter-replica TCF $\widetilde{h}(r)$ obtained from the pole-analysis-based discussions.
It is noted that the $\nu'$th mode disappears in Equation (\ref{h+ simple}) and the $\nu$th mode in Equation (\ref{h- simple});
therefore, the expressions (\ref{h+ simple}) and (\ref{h- simple}) ensure the validity of not only the real-space requirement [RR] but also the Fourier space one [FR].

The general form (\ref{h12 simple}) implies that
\begin{flalign}
\lim_{k\rightarrow k_{\nu}}\widetilde{h}(k)
&=\lim_{k\rightarrow k_{\nu}}h_+(k)
=\infty,
\label{h limit1}\\
\lim_{k\rightarrow k_{\nu'}}\widetilde{h}(k)
&=-\lim_{k\rightarrow k_{\nu'}}h_-(k)
=-\infty,
\label{h limit2}
\end{flalign}
where it is noted that the wavenumbers, $k_{\nu}$ and $k_{\nu'}$, are a subset of $k_{m}$ that represents solutions to Equation (\ref{dcf pole3}) (i.e., $\nu,\,\nu'\subset m$).
We show below that the limiting behaviors of Equations (\ref{h limit1}) and (\ref{h limit2}) can be derived from the pole Equations (\ref{dcf pole3})--(\ref{dcf pole4}) that are satisfied simultaneously at $k_{\nu}$ and $k_{\nu'}$ in accordance with the requirement [FR].

The pole Equation (\ref{main pole}), or the requirement [FR], reads
\begin{flalign}
\overline{\rho}\widetilde{h}(k_{\nu})
&=1+\overline{\rho}h(k_{\nu}),
\label{r2-1}\\
-\overline{\rho}\widetilde{h}(k_{\nu'})
&=1+\overline{\rho}h(k_{\nu'}).
\label{r2-2}
\end{flalign}
Plugging either Equation (\ref{r2-1}) or Equation (\ref{r2-2}) into Equation (\ref{oz1}), we obtain
\begin{flalign}
h(k_{\nu})&=c(k_{\nu})+\widetilde{c}(k_{\nu})+\overline{\rho}\left\{
c(k_{\nu})+\widetilde{c}(k_{\nu})\right\}h(k_{\nu}),
\label{oz second1}
\\
h(k_{\nu'})&=c(k_{\nu'})-\widetilde{c}(k_{\nu'})+\overline{\rho}\left\{
c(k_{\nu'})-\widetilde{c}(k_{\nu'})\right\}h(k_{\nu'}).
\label{oz second2}
\end{flalign}
It follows from Equations (\ref{r2-1}) to (\ref{oz second2}) that Equation (\ref{main pole}) becomes
\vspace{6pt}
\begin{flalign}
\overline{\rho}\widetilde{h}(k_{\nu})
&=\frac{1}{1-\overline{\rho}c_+(k_{\nu})},
\label{oz second3}\\
\overline{\rho}\widetilde{h}(k_{\nu'})
&=-\frac{1}{1-\overline{\rho}c_-(k_{\nu'})},
\label{oz second4}
\end{flalign}
amounting to the limiting behaviors of Equations (\ref{h limit1}) and (\ref{h limit2}) because of Equations (\ref{dcf pole3}) and (\ref{dcf pole4}) valid at $k_{\nu}$ and $k_{\nu'}$. 
Remarkably, it has been proved that $k_{\nu}$ and $k_{\nu'}$ are the poles in the liquid state.
Nevertheless, cancellation between $h_+(k)$ and $h_-(k)$ does not occur,
and we can adjust the sum of $h_+(k)$ and $h_-(k)$ to yield $h^{\mathrm{liq}}(k)$ (i.e., $h_+(k)+h_-(k)=h^{\mathrm{liq}}(k)$), according to the requirement [RR].

\begin{figure}[H]
\begin{center}
\includegraphics[
width=11cm
]{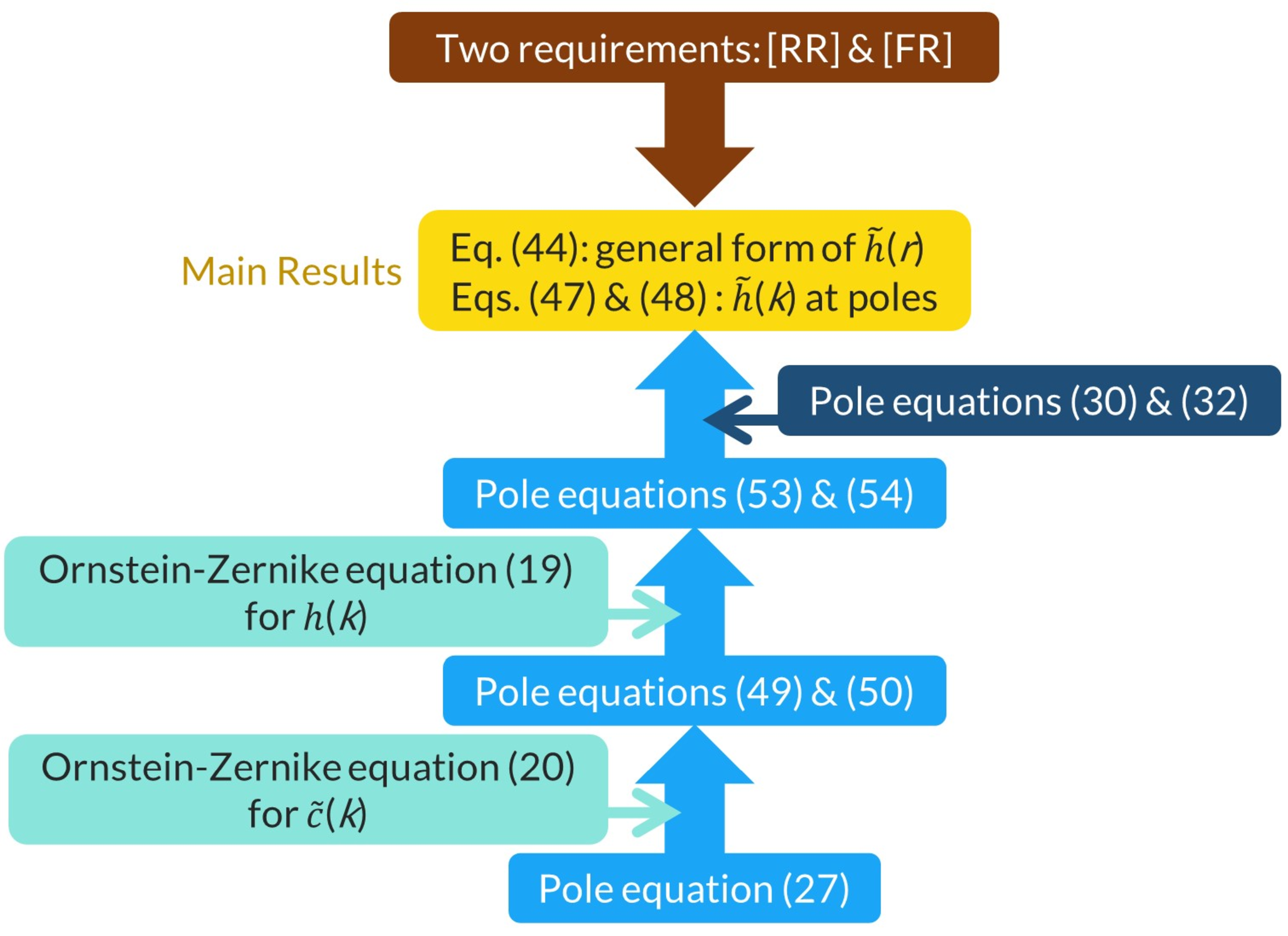}
\end{center}
\caption{A schematic summary of general formalism to obtain the main results represented by Equations~(\ref{h12 simple}) to (\ref{h limit2}).  
}\label{scheme}
\end{figure}

Figure \ref{scheme} puts together the discussions we have had so far about the two requirements, [RR] and [FR], as well as Equations (\ref{h12 simple}) to (\ref{oz second4}).
Figure \ref{scheme} clarifies the whole discussion for the general formalism, which could help us understand the overall scheme providing Equation (\ref{h12 simple}).

\subsection{Two-Mode Selection as a Minimal Model}\label{sub two mode} 
As mentioned in Section \ref{sec intro}, previous studies have confirmed that cluster glasses are characterized by two scales: cluster and cage sizes.
Focusing on these scales significant for the dynamical and structural properties, we can establish a minimal model reflecting this particularity:
we select two modes of $\nu=\mu$ and $\nu'=\mu'$, thereby reducing Equation (\ref{h12 simple}) to
\begin{flalign}
r\widetilde{h}(r)=
2A_{\mu}e^{-b_{\mu}r}\cos\left(a_{\mu}r+\theta_{\mu}\right)-2A_{\mu'}e^{-b_{\mu'}r}\cos\left(a_{\mu'}r+\theta_{\mu'}\right).
\label{h12 simple2}
\end{flalign}
Equation (\ref{h12 simple2}) implies that $b_{\mu}<b_{\mu'}$ is necessary for the first term on the rhs of \mbox{Equation~(\ref{h12 simple2})} to determine the overall behavior of $\widetilde{h}(r)$.
In other words, the decay length $b_{\mu}^{-1}$ for $p$-$p$ density fluctuations must be longer than that $b_{\mu'}^{-1}$ for $n$-$n$ density ones.

Equation (\ref{h12 simple2}) gives the inter-replica TCF at zero separation as follows:
\begin{flalign}
\widetilde{h}(0)=2A(b_{\mu'}-b_{\mu})
+2A(\delta-\delta')
\lim_{r\rightarrow 0}\frac{1}{r},
\label{zero separation}
\end{flalign}
where
\vspace{6pt}
\begin{flalign}
A_{\mu}\cos\,\theta_{\mu} =A(1+\delta),\nonumber\\
A_{\mu'}\cos\,\theta_{\mu'}=A(1+\delta'),
\label{amp}
\end{flalign}
for $b_{\mu}<b_{\mu'}$.
We see from (\ref{zero separation}) that $\widetilde{h}(0)>0$ for $\delta\geq\delta'$.
The relation $\widetilde{h}(0)>0$ indicates that there are particles of different replicas that interpenetrate each other;
$\widetilde{h}(r)$ in Equation (\ref{h12 simple2}) adequately describes the cluster glass phase where clusters are formed by molecules of two replicas, hence reaching a value of mean overlap such that $Q_{12}\gg Q_r$.

Table \ref{three modes} is another summary of our results on the two-mode model using the structure factors: $S_{\sigma}(k)$ ($\sigma=+,\,-$).
Table \ref{three modes} classifies the modes into three groups (the $\mu$th, the $\mu'$th, and the other modes) that have different characteristics in terms of density fluctuations at different wavenumbers.
Focusing on the $\mu$th mode in Table \ref{three modes}, as well as the above inequality $b_{\mu}^{-1}>b_{\mu'}^{-1}$, we see the opposite behavior as follows:
the density correlation with its oscillation period of $2\pi/a_{\mu}$ is enhanced and long-ranged in terms of $p$-density TCF $h_+(r)$, whereas the fluctuations in $n$-density are suppressed completely at the scale of $2\pi/a_{\mu}$ (i.e., $S_-(k_{\mu})=0$), according to the Fourier space requirement [FR].

\begin{table}[H]
\caption{The structure factors, $S_+(k)$, and $S_-(k)$ in the two-mode model. Three kinds of modes are characterized by different behaviors of density fluctuations at the wavenumbers of $k_{\mu}$ (the $\mu$th mode), $k_{\mu'}$ (the $\mu'$th mode), and $k_{m\neq\mu,\mu'}$ (the other modes).}
\label{three modes}
\begin{center}
\includegraphics[
width=14cm
]{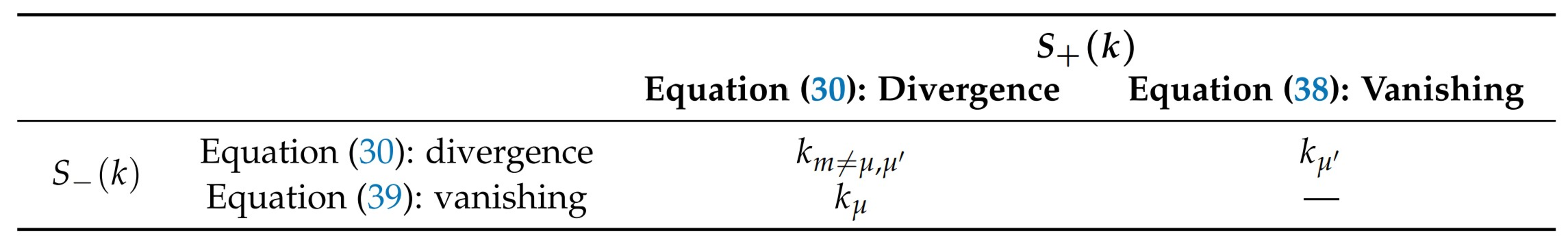}
\end{center}
\end{table}

Given a linkage between the $\mu$th mode and the cluster scale, the vanishing structure factor of $n$-density suggests the neutralization of charge within each cluster due to the complexation of two-replica clusters.
Accordingly, we find the other $\mu'$th mode related to cage size.
Section \ref{sec insights} will delve deeper into the physics underpinning the $\mu$th and $\mu'$th modes to elucidate their relationship with cluster and cage sizes with the help of FEL.
			
\section{Insights from the Free-Energy Landscape}\label{sec insights}
\begin{figure}[H]
\begin{center}
\includegraphics[
width=12cm
]{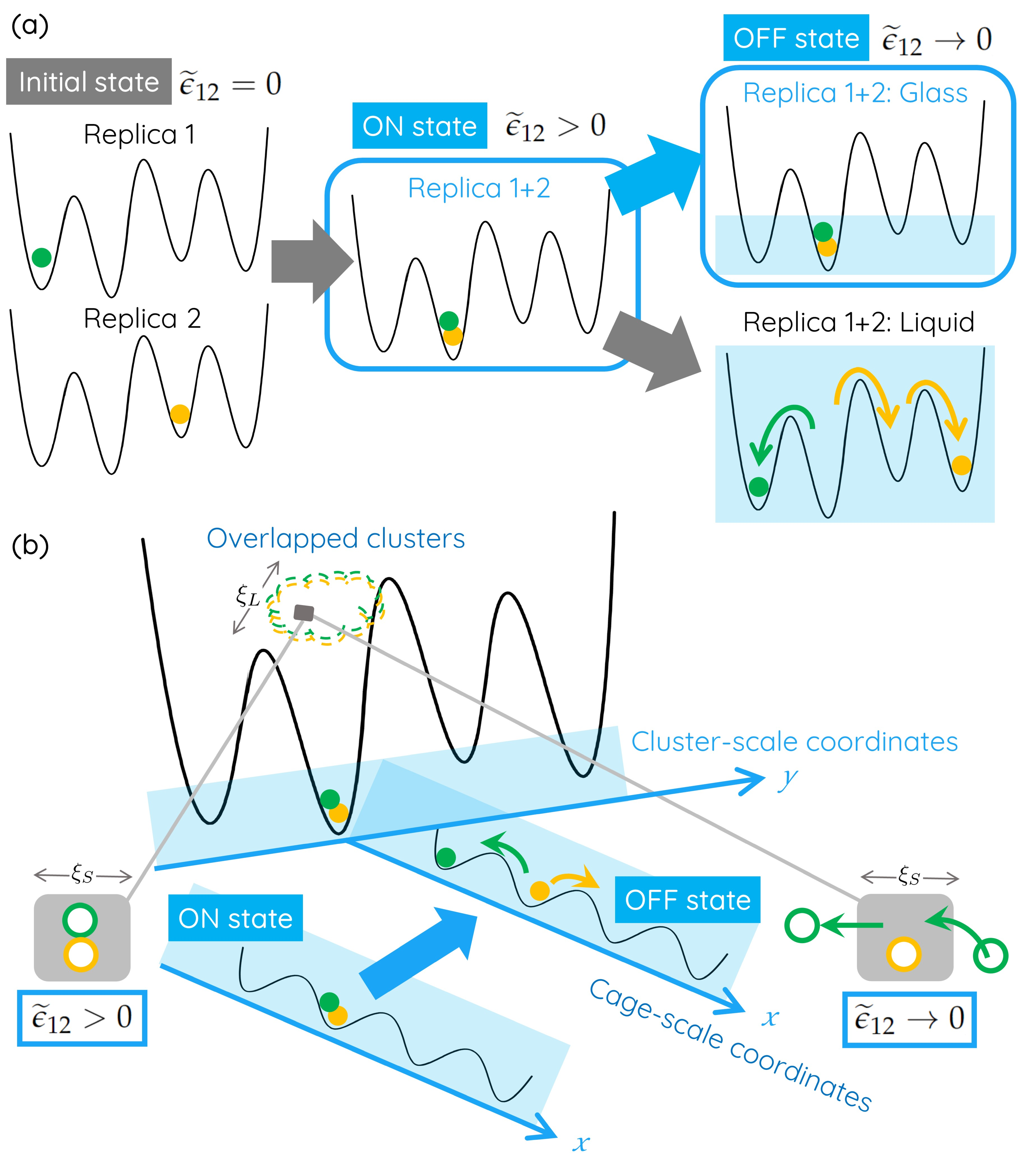}
\end{center}
\caption{Two-replica system in terms of the FEL. Green and orange circles in basins or sub-basins represent the states of different replicas.
(\textbf{a}) There are three states (i.e., Initial and ON/OFF states) when considering a couple of replicated FELs (Initial state) and the merged FEL (ON/OFF states).
Switching off the inter-replica attractive field ($\widetilde{\epsilon}_{12}\rightarrow 0$), the ON state of two replicas in the same basin transforms to the OFF state, representing either a glass or supercooled liquid.
On the one hand, two replicas trapped in the same basin represent a glass using shallow lakes separated by high mountains. On the other hand, two replicas in the liquid state can explore all configurational realizations independently because of the merging between the separated lakes.
(\textbf{b}) The bi-axial FEL for a cluster-forming glassy state.
The $x$-and $y$-axes denote particle coordinates on the scales of cage and cluster sizes (or $\xi_S$- and $\xi_L$-scales), respectively.
The FEL in the $x$-axis direction has sub-basins shallower than water depth.
In an OFF state, two replicas undergo independent transitions between the sub-basins while confined to the same basin.
}
\label{fel}
\end{figure}

Prior to assessing the validity of Equation (\ref{h12 simple2}), we reveal the underlying physics of the two-mode model in the cluster glass phase with the help of FEL.
Figure \ref{fel} exhibits different behaviors of glassy and supercooled liquid states using the FEL.
In Figure \ref{fel}a, the FEL for normal stable glasses is considered, illustrating how a single smooth basin breakes into multiple basins.
Meanwhile, the bi-axial FEL in Figure \ref{fel}b demonstrates that each basin is segmented into sub-basins.
These sub-basins represent possible configurations of cages with clusters frozen in a specific configuration (see Sections \ref{sub cluster fel} and \ref{sub two}).
The bi-axial FEL in Figure \ref{fel}b also gives insight into the roles of fluctuations in $p$-and $n$-densities, based on which we relate the requirement [FR] in Section \ref{sub req} to cluster and cage sizes: $\xi_L$ and $\xi_S$ (see Section \ref{sub two lengths}).

\subsection{The FEL for the Two-Replica System}\label{sub cluster fel}
Figure \ref{fel}a shows the change of particle configurations in the replicated FELs of two replicas, which occurs in two steps associated with switching on and off the inter-replica attractive field.
The first step is to apply the inter-replica attractive field (i.e., $-\widetilde{\epsilon}_{12}<0$) to the replicated FELs in an initial state at $\widetilde{\epsilon}_{12}=0$ of Figure \ref{fel}a.
As a consequence, we find two replicas in an identical state of the merged FEL (see an 'ON state' at $\widetilde{\epsilon}_{12}>0$ of \mbox{Figure \ref{fel}a).}
The second step is to switch off the inter-replica attractive field, creating an 'OFF state' at $\widetilde{\epsilon}_{12}\rightarrow 0$ of Figure \ref{fel}a.

The OFF state varies depending on whether the two replicas are initially in a glassy or supercooled liquid state.
The supercooled liquid state allows for two replicas to jump from the initial basin to different ones independently, thereby amounting to non-overlapped configurations.
Meanwhile, the glassy state maintains the ON state (the particle configurations in the ON state without the occurrence of transitions between adjacent basins) at $\widetilde{\epsilon}_{12}\rightarrow 0$, which is suitable for describing the cage-hopping dynamics in normal stable glasses.

However, there is a difference in the particle dynamics of normal and cluster glass states.
On the one hand, the dimer could maintain a molecular bound state in normal glasses where the cage-hopping dynamics are obviously observed;
the particles of different replicas constituting the dimer are arrested in an overlapped cage even in the absence of the attractive inter-replica field (i.e., $\widetilde{\epsilon}_{12}\rightarrow 0$).
On the other hand, we have found the blurring of cage-hopping dynamics \cite{likos g2,d1,d2,d3,d4,d5} in the cluster glass phase, indicating that the constituent particles of the dimer tend to be replaced by surrounding particles due to the loose cages.
The FEL in the cluster glass phase should be modified to reflect the above difference.

\subsection{Mixed State in Terms of the Bi-Axial FEL}\label{sub two}
In the marginal glass phase \cite{ma1,ma2,ma3,ma4,ma5,ma6,ma7}, previous theoretical studies have considered the hierarchical FEL where each basin in Figure \ref{fel}a splits into sub-basins, thereby explaining the blurring of the cage-hopping dynamics as follows: transitions between adjacent sub-basins are triggered by switching off the inter-replica attractive field while frozen in the same basin.
Considering the similarity between the particle dynamics in cluster and marginal glasses, we propose the bi-axial FEL in Figure \ref{fel}b for adding shallow sub-basins for a cluster-forming state.

While the sub-basins along the $x$-axis represent a variety of cage configurations in a specific basin, the basins along the $y$-axis present various coarse-grained arrangements of clusters.
Green and orange circles in the bi-axial FEL of Figure \ref{fel}b denote glassy states of different replicas.
These circles are located along the two axes, reflecting the hierarchical state of two-replica cluster glasses.
It is helpful to compare Figure \ref{fel}b and a set of two FELs in the rightmost column of Figure \ref{fel}a, revealing a mixed state formed by clusters in a glassy state ($y$-axis) and constituent particles in a liquid state ($x$-axis).

Let us explain the mixed state in more detail.
Focusing on the cluster scale $\xi_L$, overlapped clusters of two replicas would be indistinguishable from each other at $\xi_L$ by definition of cluster glasses.
The cluster complexation is represented by the FEL along the $y$-axis in Figure \ref{fel}b:
two replicas remain trapped at the same basin in the $y$-axis direction even after switching off the attractive interaction (i.e., $\widetilde{\epsilon}_{12}\rightarrow 0$).
Along the $x$-axis in \mbox{Figure \ref{fel}b}, on the other hand, we depict sub-basins that are shallower than water depth, thereby illustrating the ON/OFF states that are distinguishable by whether transitions between sub-basins occur or not in the $x$-axis direction.
Looking at the smaller scale of cage size $\xi_S$, a dimer formed by the inter-replica attractive field can be reconstituted in the OFF state of cluster glasses;
that is, the single-particle dynamics induce dimer rearrangements inside the overlapped cluster at $\widetilde{\epsilon}_{12}\rightarrow 0$ such that the blurring of the cage-hopping dynamics has been observed \cite{likos g2,d1,d2,d3,d4,d5}.

\subsection{Characteristic Lengths in the Two-Mode Model}\label{sub two lengths}
Let us consider the two-mode model in terms of the $p$-and $n$-density fluctuations, which vary according to the length scales (or the axes of the FEL in Figure \ref{fel}b).
The purpose here is to provide two characteristic lengths as candidates to create the two modes by revealing the underlying physics of the relations in [FR].
 
Two-replica systems exhibit two features when they belong to the same basin along the $y$-axis of Figure \ref{fel}b.
First, two replicas exhibit an identical inhomogeneity of the cluster-scale density distribution in contrast to the uniform liquid state where various configurations are possible.
Second, the molecular bound state of replicated particles in a normal glass state is replaced by the complexation of replicated clusters in the cluster glass phase.
The $p$-density, or the average density of two replicas, directly reflects the former feature.
Meanwhile, regarding the latter characteristic, the cluster complexation of two replicas leads to the disappearance of $n$-density, or no net charge, per each overlapped cluster on average.
Thus, $\xi_L$ is related to the $\mu$th mode, satisfying $S_-(k_{\mu})=0$ in [FR] as
\vspace{6pt}
\begin{flalign}
a_{\mu}=\frac{2\pi}{\xi_L}
\label{length1}
\end{flalign}
for the real part of the $\mu$th wavenumber, $k_{\mu}=a_{\mu}+ib_{\mu}$, at which there is a pole of $h_+(k)$.
Such uniformity and suppressed fluctuations in the $n$-density are in contrast to variations in cluster size and the number of particles forming each aggregate;
the polydisperse clusters necessarily produce an inhomogeneous distribution of $p$-density, $\rho_+(\bm{r})$.

Turning our attention to the other length, the cage size $\xi_S$, we can see a dimer of different replicas.
However, due to the cage-scale liquidity in the OFF state, replicas are located in different sub-basins along the $x$-axis, as illustrated in Figure \ref{fel}b.
Thus, we can suppose $\xi_S$ to be a characteristic scale over which $n$-density fluctuations, including spatiotemporal flip-flopping of charge sign, become apparent in the cluster glass phase:
\begin{flalign}
a_{\mu'}=\frac{2\pi}{\xi_S}
\label{length2}
\end{flalign}
for the real part of the $\mu'$th wavenumber, $k_{\mu'}=a_{\mu'}+ib_{\mu'}$.
Considering that the total particle number is fixed to be 2 per overlapped cage by definition, it is reasonable to satisfy \mbox{$S_+(k_{\mu'})=0$} at $k_{\mu'}=a_{\mu'}+ib_{\mu'}$ while having a pole of $h_-(k)$ at $k_{\mu'}$.

\section{Mean Field Analysis of the GCM: Assessing the Two-Mode Model}\label{sec mean}	
We assess the validity of a minimal form (\ref{h12 simple2}) based on the two-mode model.
To this end, we consider the GCM in the mean field approximation \cite{ikeda1,ikeda plus,ikeda2,coslo1,coslo2,ikeda3,miyazaki1,miyazaki2,cl1,cl2,cl3,cl4,cl5,cl6,cl7,cl8}, thereby determining not only the minimal form without the knowledge of the inter-replica DCF $\widetilde{c}(r)$ (Section \ref{sub gcm1}) but also the mode numbers, $\mu$ and $\mu'$ (Section \ref{sub gcm2}).
We compare the present results on the mean overlap $Q_{12}$ of the GCM with the previous ones at high densities where out-of-equilibrium clusters are formed (see Section \ref{sub gcm3}) \cite{likos g1,likos g2}.

\subsection{Intra-Replica TCF of the GCM in the Liquid State}\label{sub gcm1}
The two-replica system in the cluster glass phase has been numerically studied for the GCM at higher densities, showing a stepwise increase in the mean overlap $Q_{12}$:
$Q_{12}$ varies infinitesimally within a range of $\widetilde{\epsilon}$ and shows a discrete increase at a certain value of $\widetilde{\epsilon}$ \cite{likos g1}.
Hence, we investigate the GCM in the mean field approximation,
\begin{flalign}
c(k)&=-v(k)\nonumber\\  
&=-d^3\widetilde{\epsilon}\sqrt{\pi^3}
\exp\left(
-\frac{k^2}{4}\right),
\label{mf}
\end{flalign}
which has been found to be a good descriptor of the GCM at high densities \cite{likos rev1,likos rev2,ikeda1,ikeda plus,ikeda2,coslo1,coslo2,ikeda3,miyazaki1,miyazaki2,cl1,cl2,cl3,cl4,cl5,cl6,cl7,cl8}.
Equation (\ref{mf}) becomes
\begin{flalign}
\label{appendix complex gaussian}
c(k)=-d^3\widetilde{\epsilon}\sqrt{\pi^3}e^{\frac{b^2-a^2}{4}}\left\{
\cos\left(\frac{ab}{2}\right)-i\sin\left(\frac{ab}{2}\right)
\right\}
\end{flalign}
when considering the complex wavenumber $k=a+ib$ in units of $1/d$ as before.

Combining Equations (\ref{dcf pole3}) and (\ref{appendix complex gaussian}) in the above mean field approximation, the pole equation in the liquid state is expressed as
\begin{flalign}
\label{ap complex pole1}
1+\phi\,
\widetilde{\epsilon}\sqrt{\pi^3}
e^{\frac{b^2-a^2}{4}}\cos\left(\frac{ab}{2}\right)&=0,\\
\label{ap complex pole2}
\sin\left(\frac{ab}{2}\right)&=0.
\end{flalign}
Equations (\ref{ap complex pole1}) and (\ref{ap complex pole2}) reduce to
\begin{flalign}
\label{ans1}
\phi\,\widetilde{\epsilon}\sqrt{\pi^3}e^{\frac{\widetilde{b}^2-\widetilde{a}^2}{4}}&=1,\\
\label{ans2}
\frac{ab}{2}&=(2m-1)\pi,
\end{flalign}
with $m$ denoting the natural number.
The $m$th mode introduced after Equation (\ref{general tcf}) is traced back to the relation (\ref{ans2}).

Let us then consider the 3D Fourier transform of $h_{\sigma}(k)$ ($\sigma=+,\,-$) as follows:
\begin{flalign}
\label{ap fourier}
rh_{\sigma}(r)
=\frac{1}{4i\pi^2d^3}\int_{-\infty}^{\infty}dk\,k\,e^{ikr}\,h_{\sigma}(k).
\end{flalign}
We evaluate the RHS of Equation (\ref{ap fourier}) using contour integral along an infinite-radius semicircle in the complex upper half-plane.
To perform the contour integral, we need to find the poles of Equations (\ref{ap complex pole1}) and (\ref{ap complex pole2}).
Considering the four poles, $k_n^{\sigma}=a_n^{\sigma}+ib_n^{\sigma}$ and $k_{-n}^{\sigma}=-a_n^{\sigma}+ib_n^{\sigma}$ with $a_{-n}^{\sigma}=-a_n^{\sigma}$ and $b_{-n}^{\sigma}=b_n^{\sigma}$, Equation (\ref{ap fourier}) yields
\begin{flalign}
\label{ap real}
rh_{\sigma}(r)=\frac{1}{2\pi }\sum_{m\geq1}\sum_{n=m,-m}
k_n^{\sigma}e^{ik_n^{\sigma}r}\mathrm{Res}\left[h_{\sigma}(k_n^{\sigma})\right],
\end{flalign}
where the residue $\mathrm{Res}\left[h_{\sigma}(k_n^{\sigma})\right]$ of $h_{\sigma}(k_n^{\sigma})$ at a pole $k_n^{\sigma}$ reads
\begin{flalign}
\label{ap res}
\mathrm{Res}\left[h_{\sigma}(k_n^{\sigma})\right]
=\frac{-c_{\sigma}(k_n^{\sigma})}{2\phi\,c'_{\sigma}(k_n^{\sigma})},
\end{flalign}
using $c'_{\sigma}(k_n^{\sigma})=dc_{\sigma}(k)/dk|_{k=k_n^{\sigma}}$.

In the mean field approximation of Equation (\ref{mf}), we have $c'_{\sigma}(k)=-v'(k)=(k/2)v(k)$ for the GCM in the liquid state.
Accordingly, it follows from Equations (\ref{ap real}) and (\ref{ap res}), as well as the requirement [RR], that the liquid-state TCF, $h^{\mathrm{liq}}(r)=h_+(r)+h_-(r)$, in the GCM is given by
\begin{flalign}
rh^{\mathrm{liq}}(r)
&=\frac{1}{\pi}\sum_{m\geq1}\sum_{n=m,-m}
e^{-b_nr}e^{ia_nr}
\frac{k_nv(k_n)}{-2\phi\,v'(k_n)}\nonumber\\
\label{tcf gcm}
&=\frac{2}{\pi\phi}\sum_{m\geq1}
e^{-b_mr}\cos\left(a_mr\right),
\end{flalign}
implying that $A_m=1/(\pi\phi)$ and $\theta_m=0$ in Equation (\ref{eqm}).

\subsection{Two Mode Numbers Selected}\label{sub gcm2}
For comparison with previous results \cite{likos g1}, it is necessary to select the specific mode numbers of $\mu$ and $\mu'$ based on the underlying physics of [FR] for the GCM in the cluster glass phase.
Section \ref{sub two lengths} has demonstrated the relevance of the following two characteristic scales to the hierarchical dynamics in the cluster glass phase \cite{likos g2,d1,d2,d3,d4,d5}:
one length is the mean cluster diameter $\xi_L$ in Equation (\ref{length1}), whereas the other is the average size $\xi_S~(<\xi_L)$ of rattler cages in Equation (\ref{length2}).

Equations (\ref{ans1}) and (\ref{ans2}) allow us to calculate the real part of the first pole (i.e., $k_1=a_1+ib_1$).
As a result, we have $a_1\approx 5.6$ at $\phi=0.45$ and $\widetilde{\epsilon}^{-1}\times 10^{3}=1.25$, which is in good agreement with the main peak wavenumber of the intra-replica structure factor obtained numerically at the same condition in the cluster glass phase \cite{likos g1}.
Furthermore, the non-ergodic behavior of density fluctuations becomes obvious at the main peak wavenumber, which has also been demonstrated using the non-ergodicity factor of the GCM in glassy states \cite{likos g2}.
These previous results on the GCM support the validity of Equation (\ref{length1}) at $\mu=1$;
while $S_-(k)$ equals zero at the first pole $k=k_1$ of $h^{\mathrm{liq}}(k)$, the first mode represented by $k_1$ serves as a principal contribution to the $p$-$p$ density correlations, or $h_+(r)$.

Combining the above discussion with the relations (\ref{length1}) and (\ref{length2}), we find
\begin{flalign}
\mu=1,\>\mu<\mu'
\label{two modes}
\end{flalign}
as well as $\delta=\delta'=0$, ensuring $\widetilde{h}(0)>0$ in Equation (\ref{zero separation}).
It is plausible to suppose that the increase in $\widetilde{\epsilon}$ (or lowering the normalized temperature) results in the decrease in $\xi_S$ due to the cage-size reduction.
Thus, the relation (\ref{length2}) implies that the increment in the mode number $\mu'$ occurs with the gradual increase in $\widetilde{\epsilon}$.


\subsection{Comparison with Previous Results}\label{sub gcm3}
Equation (\ref{tcf gcm}) allows us to rewrite Equations (\ref{h12 simple2})--(\ref{amp}) as
\begin{flalign}
r\widetilde{h}(r)
&=\frac{2}{\pi\phi}(1+\delta)
e^{-b_\mu r}\cos\left(a_\mu r\right)
-\frac{2}{\pi\phi}(1+\delta')\,e^{-b_{\mu'} r}\cos\left(a_{\mu'} r\right),
\label{2mode gcm}\\
\widetilde{h}(0)&=\frac{2}{\pi\phi}\left(
b_{\mu'}-b_\mu
\right)
+\frac{2}{\pi\phi}
(\delta-\delta')
\lim_{r\rightarrow 0}\frac{1}{r},
\label{zero gcm}
\end{flalign}
at $\delta=\delta'=0$.

\begin{figure}[H]
\begin{center}
\includegraphics[
width=9cm
]{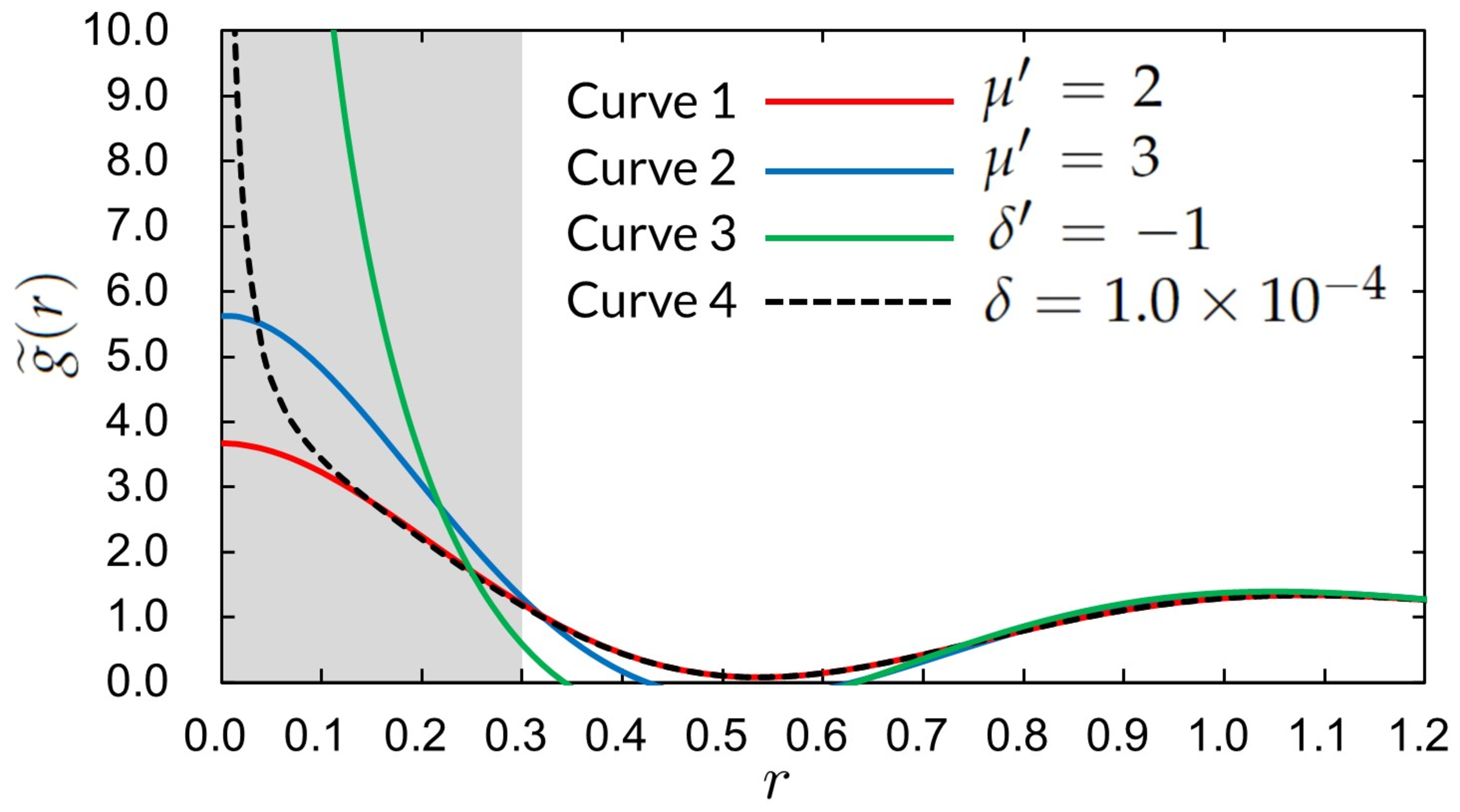}
\end{center}
\caption{All curves represent the inter-replica correlation function, $\widetilde{g}(r)=1+\widetilde{h}(r)$, given by Equation (\ref{2mode gcm}) in the mean field approximation of the GCM at $\phi=0.45$.
The solid lines show results under the same conditions as those in Table \ref{comparison}:
the red and blue lines show the profiles at $(\mu',\,\widetilde{\epsilon}^{-1}\times10^3)=(2,\,1.29)$ and $(3,\,1.25)$, respectively, and the green line is the result in the absence of the $\mu'$th mode (i.e., $\delta=0$ and $\delta'=-1$ in Equation (\ref{2mode gcm})) at $\widetilde{\epsilon}^{-1}\times10^3=1.17$.
The broken line shows a divergent behavior when making a difference such that $\delta=1.0\times 10^{-4}$ and $\delta'=0$ at $(\mu',\,\widetilde{\epsilon}^{-1}\times10^3)=(2,\,1.29)$.
The gray area indicates a range where the mean overlap $Q_{12}$ is calculated using Equations (\ref{analytic q})--(\ref{def j}).
}
\label{h of r}
\end{figure}

In Figure \ref{h of r}, there are four curves of $\widetilde{g}(r)=1+\widetilde{h}(r)$ depicted using Equations (\ref{two modes}) and (\ref{2mode gcm}).
Curve 3 depicts Equation (\ref{2mode gcm}) without the second term on the RHS, which corresponds to an approximate profile of $\widetilde{h}(r)$ in the limit of $\mu'\rightarrow\infty$.
We have finite values of~$\widetilde{h}(0)$ for curves 1 and 2 at $\delta=\delta'=0$, as implied by Equation (\ref{zero gcm}); otherwise, $\widetilde{h}(r)$ diverges at $r=0$ (see curves 3 and 4).

\vspace{-6pt}

Figure \ref{h of r} shows the following:
\begin{itemize}
\item $\widetilde{h}(0)=\widetilde{g}(0)-1>0$ is secured for all curves.
\item $\widetilde{g}(r)\geq 0$ (or $\widetilde{h}(r)\geq -1$) holds for $0\leq r \leq l$ when adopting $l=0.3$ in Equation (\ref{mean overlap}), according to the previous results \cite{likos g1}.
\item We can find a location or region of $0.3<r<1$ where either $d\widetilde{g}(r)/dr=0$ or $\widetilde{g}(r)=0$ is satisfied.
\end{itemize}
These features indicate that there is a depletion region due to the formation of a molecular cluster (or a cluster complex of two replicas) well separated from other clusters.

Figure \ref{h of r} demonstrates that we can use the expression (\ref{2mode gcm}) in Equation (\ref{meanq1}) because of $\widetilde{h}(r)\geq -1$ in the range of $0\leq r\leq l=0.3$.
Substitution of Equation (\ref{2mode gcm}) and $l=0.3$ into Equation (\ref{meanq1}) yields
\begin{flalign}
\label{analytic q}
Q_{12}-Q_r
&=8\left\{(1+\delta)I(a_1,b_1)-(1+\delta')I(a_{\mu'},b_{\mu'})\right\},\\
I(a,b)&=\int_0^{0.3}dr\,r\,e^{-br}\cos(ar)
\nonumber\\
\label{def i}
&=J(0.3)-J(0),
\end{flalign}
where
\begin{align}
\label{def j}
J(x)=\frac{\mathrm{e}^{-bx}}{(a^2+b^2)^2}\left\{2ab\sin\left(ax\right)+\left(a^2-b^2\right)\cos\left(ax\right)\right\}+\frac{x\,\mathrm{e}^{-bx}}{a^2+b^2}\left\{a\sin\left(ax\right)-b\cos\left(ax\right)\right\}.
\end{align}
We see from Equations (\ref{zero gcm}) and (\ref{analytic q}) that the change from $\delta =0$ to $\delta=1\times 10^{-4}$ at $\delta'=0$ leads to a negligible increase in $Q_{12}$, though $\widetilde{h}(0)$ diverges due to the slight shift of $\delta$.

In Table \ref{comparison}, we assess whether Equations (\ref{two modes}) to (\ref{def j}) explain the discrete increases in $Q_{12}$ observed when lowering the normalized temperature in cluster glasses of the GCM.
\mbox{Table \ref{comparison}} presents the $\widetilde{\epsilon}$-dependencies of $Q_{12}$ at three densities ($\phi=0.43$, 0.45, and 0.495) obtained from the previous numerical study \cite{likos g1}.
We calculate $Q_{12}$ at different mode numbers using Equations (\ref{analytic q})--(\ref{def j}), which are compared with the previous results in Table \ref{comparison}.
We see from Table \ref{comparison} that the discrete increase in $Q_{12}$ can be reproduced by the increase in the mode number $\mu'$ at different densities using our theory.

It is instructive to compare the profiles of solid lines (curves 1 to 3) in Figure \ref{h of r} and the mean overlap values given in Table \ref{comparison}.
On the one hand, Figure \ref{h of r} indicates that the molecular cluster is more compact and more condensed with the increase in mode number $\mu'$.
On the other hand, Table \ref{comparison} confirms that $Q_{12}$ always increases by raising $\mu'$ consistently with Equations (\ref{analytic q})--(\ref{def j}).
From combining these results, we find that the variation in $Q_{12}$ is accompanied by a sequential change as follows: the larger the degree of clusters in each replica is by lowering the temperature, the sharper the profile of the inter-replica TCF $\widetilde{h}(r)$ is around $r=0$, thereby showing a discontinuous jump in the mean overlap $Q_{12}$.

\begin{table*}
\caption{The mean overlap $Q_{12}$ of the GCM at $\phi=0.43$, 0.45 and 0.495. There are two kinds of $Q_{12}$ for each density: the upper row represents our theoretical results at the same normalized temperatures as those of previous results, whereas the lower row previous numerical results \cite{likos g1} where the values of $\widetilde{\epsilon}^{-1}\times10^3$ in parentheses correspond to normalized temperatures.
The mean values of $\widetilde{\epsilon}$ and $Q_{12}$ in each band are adopted as the previous results (see Fig. 3(a) in Ref. \cite{likos g1}).
While the theoretical results at mode numbers $\mu'=2$ to 4 are calculated using eqs. (\ref{analytic q}) to (\ref{def j}) at $\delta=\delta'=0$, our results given in the rightmost column ($\delta=0,\,\delta'=-1$) are independent of mode numbers and correspond to approximate ones in the limit of $\mu'\rightarrow\infty$ as seen from eq. (\ref{analytic q}).}
\label{comparison}
\begin{center}
\includegraphics[
width=14cm
]{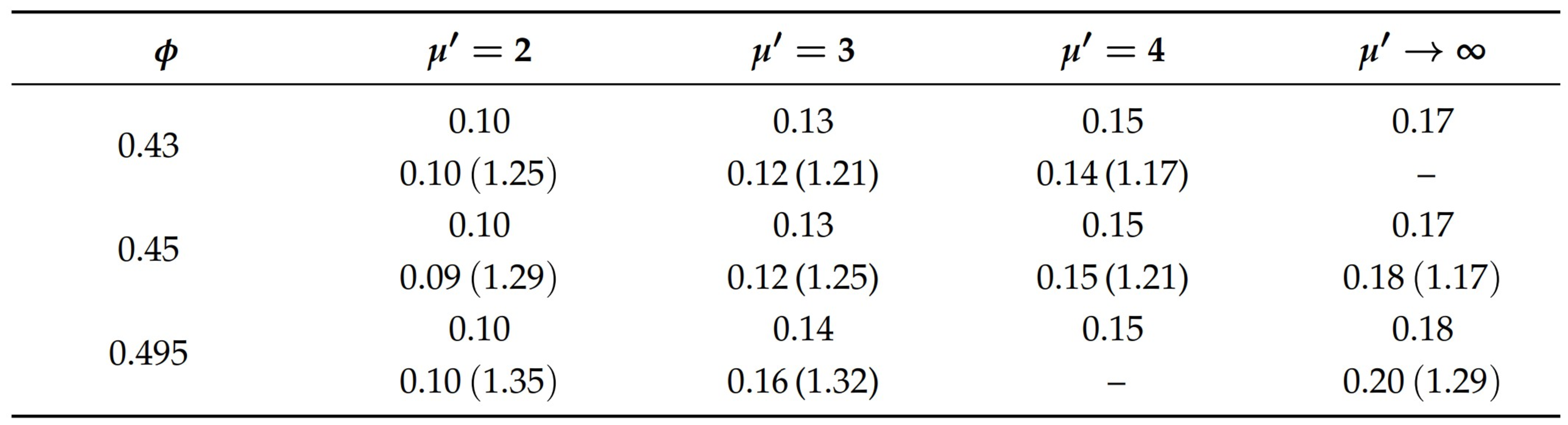}
\end{center}
\end{table*}

\section{Concluding Remarks}\label{sec concluding}
Our results consist of three parts.
Each part presents the pole-analysis-based expression for the inter-replica TCF $\widetilde{h}(r)$, which becomes simple in the later part.
(i) The first part (Sections \ref{sec general} and \ref{sub general}) has presented two methods by which to obtain a general form (\ref{h12 simple}) from the pole analysis.
(ii) In the second part (Sections \ref{sub two mode} and \ref{sec insights}), we have introduced the two-mode model represented by Equation (\ref{h12 simple2}). 
The second part has further related two characteristic lengths ($\xi_L$ and $\xi_S$) to the two modes in Equations (\ref{length1}) and (\ref{length2}), respectively, based on insights from the bi-axial FEL (Figure \ref{fel}) as an illustration of the hierarchical dynamics inherent in cluster glasses.
(iii) The last part (Section \ref{sec mean}) has applied the two-mode model to the GCM at high densities, thereby leading to the most simplified form (\ref{2mode gcm}).
Remarkably, Equation (\ref{2mode gcm}) yields Equations (\ref{analytic q})--(\ref{def j}), the mean overlap $Q_{12}$ obtained in the mean field approximation, which requires no fitting parameters.
Table \ref{comparison} shows that the analytical expression (\ref{analytic q}) for $Q_{12}$ gives theoretical results that are in good agreement with those from the previous numerical study \cite{likos g1}.

To be clear in our conclusion, we provide additional notes on each part:
\begin{enumerate}
\item[(i)] The flow to obtain Equation (\ref{h12 simple}) is summarized in Figure \ref{scheme}, illustrating that two approaches lead to the same form, Equation (\ref{h12 simple}). 
While the validity of Equation (\ref{h12 simple}) is ensured by two physical requirements ([RR] and [FR]), we can derive Equation (\ref{h12 simple}) from combining the OZ equations (Equations (\ref{oz1}) and (\ref{oz2})) and the pole equation (Equation (\ref{dcf pole})).
It is also important to note that spatial isotropy is assumed for TCFs, $h(r)$ and $\widetilde{h}(r)$, in the OZ equations.
Applying this supposition to cluster glasses is equivalent to assuming the isotropic shape of each cluster.
In other words, the morphological diversity of clusters is beyond the scope of this study, taking no account of the surface energy of clusters.
\item[(ii)] Figure \ref{fel} illustrates the shift from ON to OFF state associated with switching off the attractive field (i.e., $\widetilde{\epsilon}_{12}\rightarrow 0$) when considering the FEL for the two-replica system (i.e., $n=2$).
The applied attractive field allows for two replicas to select the same basin in the $y$-axis direction of Figure \ref{fel}b.
After switching off the attractive field, two replicas in the cluster glass phase remain trapped in the same basin but are no longer confined within the same sub-basin in the $x$-axis direction of Figure \ref{fel}b, which causes the blurring of cage-hopping dynamics.
\item[(iii)] For assessing the analytical theory, we focus on discrete increases in $Q_{12}$ due to lowering the normalized temperature $\widetilde{\epsilon}^{-1}$ in cluster glasses of the GCM.
We can ascribe the discontinuous change to the second term on the RHS of Equation (\ref{2mode gcm}) that arises from the $\mu'$th mode of $h_-(r)$ defined in Equation (\ref{n tcf}).
Surprisingly, Table \ref{comparison} supports the validity of our theory by demonstrating that the novel $\widetilde{\epsilon}$-dependencies of $Q_{12}$ can be explained from the mode number $\mu'$ that increases by one.
\end{enumerate}

Finally, going back to the general form (\ref{h12 simple}) of the inter-replica TCF, we would like to see what is implied by the difference between the first and second terms on the RHS of Equation (\ref{h12 simple}).
On the one hand, the first term with a positive sign in Equation (\ref{h12 simple}) expresses synchronous cluster-scale fluctuations of two replicas confined to the same basin (i.e., fluctuations along the $y$-axis in Figure \ref{fel}b).
On the other hand, the second term with a negative sign in Equation (\ref{h12 simple}) represents fluctuations along the $x$-axis in Figure \ref{fel}b, causing a decrease in inter-replica correlations due to asynchronous behaviors of different replica particles such as vibration around different sub-basins and transition to other sub-basins.

The theoretical development would help us understand the underlying physics of cluster glasses. First, recall that the GCM discussed in Section \ref{sec mean} corresponds to a standard model of polymeric systems \cite{likos rev1,likos rev2}. Additionally, colloidal systems with competing interactions (short-range attractive and long-range repulsive interactions) as well as the $Q^{\pm}$ class in the GEM vitrify while forming equilibrium clusters with diverse morphologies \cite{com1,com7}. For any of these systems, the pole-analysis-based discussions are valid. Specifically, the two-mode model, which focuses on two lengths (cluster and particle scales), is suitable for investigating the FELs in cluster glasses as detailed in Section \ref{sec insights}. Our two-replica theory can also address the dynamical heterogeneity in realistic glassy systems because the overlap correlation function provides the four-point correlation function, a significant measure of dynamic heterogeneity \cite{add1,add2}. Therefore, our simple method of evaluating overlap could offer a promising approach to investigate dynamical issues in glassy systems with hierarchical structures, including cluster glasses. However, it remains necessary to compare experimental and simulation results for various glassy systems with our theoretical ones obtained from more accurate forms of the DCFs, which will allow us to gain deeper insights into the cluster glass phase.

\subsection*{Nomenclature}
\begin{center}
\begin{tabular}{@{}ll}
GEM & generalized exponential model\\
GCM & Gaussian core model\\
FEL & free-energy landscape\\
TCF & total correlation function\\
DCF & direct correlation function\\
OZ equation & Ornstein--Zernike equation\\
$v(r)$ & interaction potential in units of $k_BT$ \\
$r$ & interparticle distance in units of a characteristic length $d$ \\
$\widetilde{\epsilon}$ & dimensionless interaction strength at zero separation \\
$\widehat{\rho}_{\alpha}(\bm{x})$ & microscopic density in replica $\alpha$ of $N$-particle system  \\
$Q_{\alpha\beta}$ & mean overlap between configurations of replica $\alpha$ and replica $\beta$\\
$\phi=\overline{\rho}d^3$ & volume fraction defined using a uniform density $\overline{\rho}$\\
$\widehat{\rho}_+(\bm{r})$ & $p$-density defined in Equation (\ref{plus density})\\
$\widehat{\rho}_-(\bm{r})$ & $n$-density defined in Equation (\ref{minus density})\\
$h_+(r)$ & TCF of $p$-$p$ density fluctuations defined in Equation (\ref{p total})\\
$h_-(r)$ & TCF of $n$-$n$ density fluctuations defined in Equation (\ref{n total})\\
$h(r)$ & intra-replica TCF\\
$\widetilde{h}(r)$ & inter-replica TCF\\
$c(r)$ & intra-replica DCF\\
$\widetilde{c}(r)$ & inter-replica DCF\\
$S_+(k)$ & structure factor of $p$-$p$ density fluctuations defined in Equation (\ref{s relation2})\\
$S_-(k)$ & structure factor of $n$-$n$ density fluctuations defined in Equation (\ref{s relation2})\\
\end{tabular}     
\end{center}
\newpage


\bibliographystyle{apsrev4-1}

\begin{thebibliography}{999}
\bibitem{likos rev1}
Likos, C.N. Effective interactions in soft condensed matter physics. {\itshape Phys. Rep.} {\bf 2001}, \emph{348}, 267--439.
\bibitem{likos rev2}
Likos, C.N. Soft matter with soft particles. {\itshape Soft Matter} {\bf 2006}, \emph{2}, 478--498
\bibitem{van hecke}
van Hecke, M. Jamming of soft particles: Geometry, mechanics, scaling and isostaticity. {\itshape J. Phys. Condens. Matter} {\bf 2009}, \emph{22}, 033101.
\bibitem{nature}
Cinti, F.; Macr\'i, T.; Lechner, W.; Pupillo, G.; Pohl, T. Defect-induced supersolidity with soft-core bosons. {\itshape Nat. Commun.} {\bf 2014}, \emph{5}, 3235.
\bibitem{vortex}
D\'iaz-M\'endez, R.; Mezzacapo, F.; Lechner, W.; Cinti, F.; Babaev, E.; Pupillo, G. Glass transitions in monodisperse cluster-forming ensembles: Vortex matter in type-1.5 superconductors. {\itshape Phys. Rev. Lett.} {\bf 2017}, \emph{118}, 067001.

\bibitem{ikeda1}
Ikeda, A. Miyazaki, K. Glass transition of the monodisperse Gaussian core model. {\itshape Phys. Rev. Lett.} {\bf 2011}, \emph{106}, 015701.
\bibitem{ikeda plus}
Ikeda, A.; Miyazaki, K. Thermodynamic and structural properties of the high density Gaussian core model. {\itshape J. Chem. Phys.} {\bf 2011}, \emph{135}, 024901.
\bibitem{ikeda2}
Ikeda, A. Miyazaki, K. Slow dynamics of the high density gaussian core model. {\itshape J. Chem. Phys.} {\bf 2011}, \emph{135}, 054901.
\bibitem{coslo1}
Coslovich, D. Bernabei, M. Moreno, A.J. Cluster glasses of ultrasoft particles. {\itshape J. Chem. Phys.} {\bf 2012}, \emph{137}, 184904.
\bibitem{coslo2}
Coslovich, D. Ikeda, A. Cluster and reentrant anomalies of nearly Gaussian core particles. {\itshape Soft Matter} {\bf 2013}, \emph{9}, 6786--6795.
\bibitem{ikeda3}
Coslovich, D. Ikeda, A. Miyazaki, K. Mean-field dynamic criticality and geometric transition in the Gaussian core model. {\itshape Phys. Rev. E} {\bf 2016}, \emph{93}, 042602.
\bibitem{miyazaki1}
Miyazaki, R.; Kawasaki, T.; Miyazaki, K. Cluster glass transition of ultrasoft-potential fluids at high density. {\itshape Phys. Rev. Lett.} {\bf 2016}, \emph{117}, 165701.
\bibitem{miyazaki2}
Miyazaki, R.; Kawasaki, T.; Miyazaki, K. Slow dynamics coupled with cluster formation in ultrasoft-potential glasses. {\itshape J. Chem. Phys.} {\bf 2019}, \emph{150}, 074503.

\bibitem{cl1}
Louis, A.A.; Bolhuis, P.G.; Hansen, J.P. Mean-field fluid behavior of the Gaussian core model. {\itshape Phys. Rev. E} {\bf 2000}, \emph{62}, 7961.
\bibitem{cl2}
Lang, A.; Likos, C.N.; Watzlawek, M.; L\"owen, H. Fluid and solid phases of the Gaussian core model. {\itshape J. Phys. Condens. Matter} {\bf 2000}, \emph{12}, 5087.
\bibitem{cl3}
Likos, C.N.; Lang, A.; Watzlawek, M.; L\"owen, H. Criterion for determining clustering versus reentrant melting behavior for bounded interaction potentials. {\itshape Phys. Rev. E} {\bf 2001}, \emph{63}, 031206.
\bibitem{cl4}
Mladek, B.M.; Gottwald, D.; Kahl, G.; Neumann, M.; Likos, C.N. Formation of polymorphic cluster phases for a class of models of purely repulsive soft spheres. {\itshape Phys. Rev. Lett.} {\bf 2006}, \emph{96}, 045701.
\bibitem{cl5}
Mladek, B.M.; Gottwald, D.; Kahl, G.; Neumann, M.; Likos, C.N. Clustering in the absence of attractions: Density functional theory and computer simulations. {\itshape J. Phys. Chem. B} {\bf 2007}, \emph{111}, 12799--12808.
\bibitem{cl6}
Likos, C.N.; Mladek, B.M.; Gottwald, D.; Kahl, G. Why do ultrasoft repulsive particles cluster and crystallize? Analytical results from density-functional theory. {\itshape J. Chem. Phys.} {\bf 2007}, \emph{126}, 224502.
\bibitem{cl7}
Pini, D.; Parola, A.; Reatto, L. An unconstrained DFT approach to microphase formation and application to binary Gaussian mixtures. {\itshape J. Chem. Phys.} {\bf 2015}, \emph{143}, 034902.
\bibitem{cl8}
Nikiteas, I.; Heyes, D.M. Reentrant melting and multiple occupancy crystals of bounded potentials: Simple theory and direct observation by molecular dynamics simulations. {\itshape Phys. Rev. E} {\bf 2020}, \emph{102}, 042102.

\bibitem{likos g1}
Bomont, J.M.; Likos, C.N.; Hansen, J.P. Glass quantization of the Gaussian core model. {\itshape Phys. Rev. E} {\bf 2022}, \emph{105}, 024607.
\bibitem{likos g2}
Sposini, V.; Likos, C.N.; Camargo, M. Glassy phases of the Gaussian core model. {\itshape Soft Matter} {\bf 2023}, \emph{19}, 9531--9540.

\bibitem{d1}
Montes-Saralegui, M.; Nikoubashman, A.; Kahl, G. Merging and hopping processes in systems of ultrasoft, cluster forming particles under compression. {\itshape J. Chem. Phys.} {\bf 2014}, \emph{141}, 124908.
\bibitem{d2}
Schwanzer, D.F.; Coslovich, D.; Kahl, G. Two-dimensional systems with competing interactions: Dynamic properties of single particles, and of clusters. {\itshape J. Phys. Condens. Matter} {\bf 2016}, \emph{28}, 414015.
\bibitem{d3}
Liu, Y.; Liu, G.; Zhang, W.; Du, C.; Wesdemiotis, C.; Cheng, S.Z. Cooperative soft-cluster glass in giant molecular clusters. {\itshape Macromolecules} {\bf 2019}, \emph{52}, 4341--4348.
\bibitem{d4}
Cho, J.H.; Cerbino, R.; Bischofberger, I. Emergence of multiscale dynamics in colloidal gels. {\itshape Phys. Rev. Lett.} {\bf 2020}, \emph{124}, 088005.
\bibitem{d5}
Liebetreu, M.; Likos, C.N. Shear-induced stack orientation, and breakup in cluster glasses of ring polymers. {\itshape ACS Appl. Polym. Mater.} {\bf 2020}, \emph{2}, 3505--3517.

\bibitem{ma1}
Charbonneau, P.; Kurchan, J.; Parisi, G.; Urbani, P.; Zamponi, F. Fractal free energy landscapes in structural glasses. {\itshape Nat. Commun.} {\bf 2014}, \emph{5}, 3725.
\bibitem{ma2}
M\"uller, M.; Wyart, M. Marginal stability in structural, spin, and electron glasses. {\itshape Annu. Rev. Condens. Matter Phys.} {\bf 2015}, \emph{6}, 177--200.
\bibitem{ma3}
Berthier, L.; Biroli, G.; Charbonneau, P.; Corwin, E.I.; Franz, S.; Zamponi, F. Gardner physics in amorphous solids and beyond. {\itshape \mbox{J. Chem. Phys.}} {\bf 2019}, \emph{151}, 010901.
\bibitem{ma4}
Dennis R.C.; Corwin, E.I. Jamming energy landscape is hierarchical and ultrametric. {\itshape Phys. Rev. Lett.} {\bf 2020}, \emph{124}, 078002.
\bibitem{ma5}
Hammond, A.P.; Corwin, E.I. Experimental observation of the marginal glass phase in a colloidal glass. {\itshape Proc. Natl. Acad. Sci. USA} {\bf 2020}, \emph{117}, 5714--5718.
\bibitem{ma6}
Ikeda, H.; Miyazaki, K.; Yoshino, H.; Ikeda, A. Multiple glass transitions and higher-order replica symmetry breaking of binary mixtures. {\itshape Phys. Rev. E} {\bf 2021}, \emph{103}, 022613.
\bibitem{ma7}
Arceri, F.; Corwin, E.I.; O'Hern, C.S. The jamming transition and the marginally stable solid. In {\itshape Spin Glass Theory and Far Beyond: Replica Symmetry Breaking after 40 Years}; Marinari, E., M\'ezard, M., Parisi, G., Ricci-Tersenghi, F., Sicuro, G., Zamponi, F., Eds.; World Scientific: Singapore, 2023; pp. 239--254.

\bibitem{qu1}
Franz, S.; Parisi, G. Phase diagram of coupled glassy systems: A mean-field study. {\itshape Phys. Rev. Lett.} {\bf 1997}, \emph{79}, 2486--2489.
\bibitem{qu2}
Franz, S.; Parisi, G. Effective potential in glassy systems: Theory and simulations. {\itshape Phys. A Stat. Mech.} {\bf 1998}, \emph{261}, 317--339.
\bibitem{qu3}
Cardenas, M.; Franz, S.; Parisi, G. Constrained Boltzmann-Gibbs measures and effective potential for glasses in hypernetted chain approximation and numerical simulations. {\itshape J. Chem. Phys.} {\bf 1999}, \emph{110}, 1726--1734.
\bibitem{qu4}
Franz, S.; Parisi, G. On non-linear susceptibility in supercooled liquids. {\itshape J. Phys. Condens. Matter} {\bf 2000}, \emph{12}, 6335.
\bibitem{qu5}
Jack, R.L.; Berthier, L. The melting of stable glasses is governed by nucleation-and-growth dynamics. {\itshape J. Chem. Phys.} {\bf 2016}, \emph{144}, 244506.
\bibitem{qu6}
Guiselin, B.; Berthier, L. Tarjus, G. Statistical mechanics of coupled supercooled liquids in finite dimensions. {\itshape SciPost Phys.} {\bf 2022}, \emph{12}, 091.
\bibitem{qu7}
Frusawa, H. Replica Field Theory for a Generalized Franz–Parisi Potential of Inhomogeneous Glassy Systems: New Closure and the Associated Self-Consistent Equation. {\itshape Entropy} {\bf 2024}, \emph{26}, 241.

\bibitem{an1}
Berthier, L. Overlap fluctuations in glass-forming liquids. {\itshape Phys. Rev. E} {\bf 2013}, \emph{88}, 022313.
\bibitem{an2}
Parisi, G.; Seoane, B. Liquid-glass transition in equilibrium. {\itshape Phys. Rev. E} {\bf 2014}, \emph{89}, 022309.
\bibitem{an3}
Bomont, J.M.; Hansen, J.P.; Pastore, G. An investigation of the liquid to glass transition using integral equations for the pair structure of coupled replicae. {\itshape J. Chem. Phys.} {\bf 2014}, \emph{141}, 174505.
\bibitem{an4}
Bomont, J.M.; Hansen, J.P.; Pastore, G. Hypernetted-chain investigation of the random first-order transition of a Lennard-Jones liquid to an ideal glass. {\itshape Phys. Rev. E} {\bf 2015}, \emph{92}, 042316.
\bibitem{an5}
Bomont, J.M.; Pastore, G. An alternative scheme to find glass state solutions using integral equation theory for the pair structure. {\itshape Mol. Phys.} {\bf 2015}, \emph{113}, 2770--2775.
\bibitem{an6}
Bomont, J.M.; Hansen, J.P.; Pastore, G. Revisiting the replica theory of the liquid to ideal glass transition. {\itshape J. Chem. Phys.} {\bf 2019}, \emph{150}, 154504.
\bibitem{an7}
Bomont, J.M.; Pastore, G.; Hansen, J.P. Coexistence of low and high overlap phases in a supercooled liquid: An integral equation investigation. {\itshape J.Chem. Phys.} {\bf 2019}, \emph{146}, 114504.

\bibitem{evans1}
Evans, R.; Leote de Carvalho, R.J.F.; Henderson, J.R.; Hoyle, D.C. Asymptotic decay of correlations in liquids and their mixtures. {\itshape J. Chem. Phys.} {\bf 1994}, \emph{100}, 591--603.
\bibitem{evans2}
Dijkstra, M.; Evans, R. A simulation study of the decay of the pair correlation function in simple fluids. {\itshape J. Chem. Phys.} {\bf 2000}, \emph{112}, 1449--1456.
\bibitem{evans3}
Grodon, C.; Dijkstra, M.; Evans, R.; Roth, R. Decay of correlation functions in hard-sphere mixtures: Structural crossover. {\itshape J. Chem. Phys.} {\bf 2004}, \emph{{121}}, 7869--7882.
\bibitem{evans4}
Archer, A.J.; Pini, D.; Evans, R.; Reatto, L. Model colloidal fluid with competing interactions: Bulk and interfacial properties. {\itshape \mbox{J. Chem. Phys.}} {\bf 2007}, \emph{126}, 014104.
\bibitem{evans5}
Walters, M.C.; Subramanian, P.; Archer, A.J.; Evans, R. Structural crossover in a model fluid exhibiting two length scales: Repercussions for quasicrystal formation. {\itshape Phys. Rev. E} {\bf 2018}, \emph{98}, 012606.
\bibitem{evans6}
Cats, P.; Evans, R.; H\"artel, A.; van Roij, R. Primitive model electrolytes in the near, and far field: Decay lengths from DFT, and simulations. {\itshape J. Chem. Phys.} {\bf 2021}, \emph{154}, 124504.

\bibitem{com1}
Klix, C.L.; Royall, C.P.; Tanaka, H. Structural and dynamical features of multiple metastable glassy states in a colloidal system with competing interactions. {\itshape Phys. Rev. Lett.} {\bf 2010}, \emph{104}, 165702.
\bibitem{com2}
Tsurusawa, H.; Leocmach, M.; Russo, J.; Tanaka, H. Direct link between mechanical stability in gels and percolation of isostatic particles. {\itshape Sci. Adv.} {\bf 2019}, \emph{5}, eaav6090.
\bibitem{com3}
Ruiz-Franco, J.; Zaccarelli, E. On the role of competing interactions in charged colloids with short-range attraction. {\itshape Annu. Rev. Condens. Matter Phys.} {\bf 2021}, \emph{12}, 51--70.
\bibitem{com4}
Tan, J.; Afify, N.D.; Ferreiro-Rangel, C.A.; Fan, X.; Sweatman, M.B. Cluster formation in symmetric binary SALR mixtures. {\itshape J. Chem. Phys.} {\bf 2021}, \emph{154}, 074504.
\bibitem{com5}
Costa, D.; Muna\'o, G.; Bomont, J.M.; Malescio, G.; Palatella, A.; Prestipino, S. Microphase versus macrophase separation in the square-well-linear fluid: A theoretical and computational study. {\itshape Phys. Rev. E} {\bf 2023}, \emph{108}, 034602.
\bibitem{com6}
Patsahan, O.; Meyra, A.; Ciach, A. Spontaneous pattern formation in monolayers of binary mixtures with competing interactions. {\itshape Soft Matter} {\bf 2024}, \emph{20}, 1410--1424.
\bibitem{com7}
Bomont, J.M.; Pastore, G.; Costa, D.; Muna\'o, G.; Malescio, G.; Prestipino, S. Arrested states in colloidal fluids with competing interactions: A static replica study. {\itshape J. Chem. Phys.} {\bf 2024}, \emph{160}, 214504.

\bibitem{add1}
Franz, S.; Jacquin, H.; Parisi, G.; Urbani, P.; Zamponi, F. Static replica approach to critical correlations in glassy systems. {\itshape J. Chem. Phys.} {\bf 2013}, \emph{138}, 12A540.
\bibitem{add2}
Folena, G.; Biroli, G.; Charbonneau, P.; Hu, Y.; Zamponi, F. Equilibrium fluctuations in mean-field disordered models. {\itshape Phys. Rev. E} {\bf 2022}, \emph{106}, 024605.
\end{thebibliography}

\end{document}